\documentclass[useAMS]{mn2e}
\usepackage{graphicx}
\begin{document}
\def\simlt{\mathrel{\rlap{\lower 3pt\hbox{$\sim$}}
        \raise 2.0pt\hbox{$<$}}}
\def\simgt{\mathrel{\rlap{\lower 3pt\hbox{$\sim$}}
        \raise 2.0pt\hbox{$>$}}}
\def\bj{b_{\rm\scriptscriptstyle J}}
\def\rt{r_{\rm\scriptscriptstyle T}}
\def\rp{r_{\rm\scriptscriptstyle P}}
\renewcommand{\labelenumi}{(\arabic{enumi})}
\title[The interplay between radio galaxies and cluster environment]
{The interplay between radio galaxies and cluster environment}
\author[M. Magliocchetti \& M. Br\"uggen]
{\parbox[t]\textwidth{Manuela Magliocchetti$^{1,2,3}$ \& Marcus 
Br\"uggen$^4$ }\\
\tt $^1$ INAF, Osservatorio Astronomico di Trieste, Via Tiepolo 11, 34100,
Trieste, Italy\\
\tt $^2$ ESO, Karl-Schwarzschild-Str.2, D-85748, Garching, Germany\\
\tt $^3$ S.I.S.S.A., via Beirut 2-4, 34100, Trieste, Italy\\
\tt $^4$ Jacobs University Bremen, Campus Ring 1, 28759 Bremen, Germany}

\maketitle

\vspace{7cm}
\begin{abstract}
%to be rechecked
By combining the REFLEX and NORAS cluster datasets with the NVSS radio 
catalogue, we obtain a sample of 145, $z<0.3$, X-ray selected clusters brighter than 
$3\cdot 10^{-12}$~erg s$^{-1}$cm$^{-2}$ that show a central radio emission above 
3 mJy. For virial masses $M_{\rm vir}\simlt 10^{14.5} M_\odot$, 11 clusters out of 12 
(corresponding to 92\% of the systems) are inhabited by a central radio source. 
This fraction decreases with higher masses as $\propto M_{\rm vir}^{-0.4}$. 
If this decrease is a selection effect, it
%Although due to a selection effect, this result 
suggests that the majority of X-ray selected clusters host in their centre a 
radio source brighter than $\sim 10^{20}$ W/Hz/sr.  
A division of the sample into clusters harbouring either point-like or 
an extended radio-loud AGN reveals that the steepening of the 
$L_X-T$ relation for low-temperature clusters is strongly associated with the presence 
of central radio objects with extended jets and/or lobe structures. 
In the latter case, $L_X\propto T^{4}$ while for point-like sources one recovers an 
approximately self-similar relation $L_X\propto T^{2.3}$. Monte Carlo simulations 
show that the steepening of the $L_X-T$ relation is not caused by 
clusters being under-luminous in the X-ray band, but rather by overheating, 
most likely caused by the interplay between the extended radio structures  
and the intracluster medium. In the case of low-mass 
systems, we also find a tight correlation between radio luminosity and 
cluster temperature.
The effects of the central radio source on the thermal state of 
a cluster become less important with increasing cluster mass.\\
%contribution of AGN heating from a central radio source however becomes 
%less important in the case of high mass systems. 
The presence of radio sources with extended structures 
(61, corresponding to $\sim 42\%$ of the sample) is enhanced 
in X-ray luminous clusters with respect to 'field' radio-loud AGN.  
Furthermore, we find that the luminosity distribution of the cluster radio population 
differs from that of all radio sources, 
as there is a deficit of low-luminosity ($L_R\simlt 10^{22}$~W/Hz/sr) objects, while  
the number of high-luminosity ones is boosted. 
The net effect on the radio luminosity function of radio galaxies 
associated to cluster centres is of a flattening at all luminosities 
$L_R\simlt 10^{24}$~W/Hz/sr. 
%by favouring high-luminosity objects ($L_R\simgt 10^{22}$~W/Hz/sr).

%The result that the effects of 
%radio-source induced heating on the thermal properties of clusters are strongest 
%for extended radio structures suggests that AGN heating is a short-term effect 
%associated with recent AGN activity.

%for (i) low-mass systems and (ii) for extended radio structures.   

\end{abstract}

\begin{keywords}
cosmology:observations - galaxies:clusters:general - galaxies:active -  
galaxies:statistics-  X-rays:galaxies:clusters - radio continuum:galaxies
\end{keywords}

%%%%%%%%%%%%%%%%%%%%%%%%%%%%%%%%%%%%%%%%%%%%%%%%%%%%%%%%%%%%%%%%%%%%%%%%%%%%%%%%%%%%%

\section {Introduction}

Arguably one of the most important problems in the field of structure formation is the 
question of how cooling of gas, which leads to star and galaxy formation, is controlled 
by various heating processes.
The widely accepted, hierarchical picture for galaxy formation states that baryonic matter 
accretes onto dark matter halos. This picture implies that in the absence of any 
feedback process, the galaxy mass function follows the dark matter halo mass function. 
However, it has been known for some time that this is not true (see e.g. White \& Frenk 1991). 
At the high- and low-mass end, there is a deficit of galaxies compared with dark matter 
halos. These discrepancies indicate that various heating mechanisms must be involved 
in the formation of galaxies. Galaxy clusters possess the largest dark matter halos found 
in the universe and contain gas at X-ray-emitting temperatures. The intracluster medium (ICM) 
of galaxy clusters 
has been studied intensively and X-ray selected clusters of galaxies provide an excellent 
laboratory to explore the complex interplay between gas properties and heating mechanisms.\\

Galaxy clusters are known to obey fairly tight scaling relations between global properties 
such as mass, temperature and X-ray luminosity. These relations are important when 
using galaxy clusters to estimate cosmological parameters. 
It is also known that simple, self-similar models (Kaiser et al. 
1986) that correctly reproduce the scaling properties of the dark-matter distribution 
of galaxy clusters (e.g. Borgani et al. 2001; Pointecouteau et al. 2005; Vikhlinin et 
al. 2006) fail in describing the observed scaling relations of the hot baryonic component.
This discrepancy increases as one moves from 
the regime of rich clusters of galaxies to that of poor clusters and groups, giving rise to 
the so-called entropy-floor problem (e.g. Ponman et al. 1999; Ponman et al. 2003; 
Pratt \& Arnaud 2005; Piffaretti et al. 2005). Some source of extra heating has to be 
invoked to reconcile theoretical expectations and observations.\\
 
Moreover, in the absence of non-gravitational heating, very dense cluster cores in which 
the cooling time is often less than the Hubble time (so-called cool core clusters) should 
cool and accrete gas at rates of hundreds and more solar masses per year (see Fabian et al. 
2004 for a review). However, this model conflicts with results from X-ray spectroscopy that 
have provided evidence that there is no massive cooling of gas in the central regions of 
cool core clusters. (e.g. Peterson et al. 2001; 2003; B\"ohringer et al. 2002). Numerical 
simulations of clusters of galaxies that include radiative cooling (e.g. Bryan 2000; Voit 
\& Brian 2001; Wu \& Xue 2002) 
star formation and supernova feedbacks (e.g. Borgani et al. 2004; 2005) are found to be 
unable to reproduce the observed properties of the intracluster medium.\\
%Energy injections of various forms at high redshifts 
%(e.g. Evrard \& Henry 1991; Kaiser 1991; Tozzi \& Norman 2001; Dos Santos \& Dore' 2002) 
%and composite models which allow for some form of pre-heating (e.g. Kay, Thomas \& 
%Theuns 2003; %Tornatore et al. 2003; Borgani et al. 2005) seem to do a better job

Currently, the most popular model that is invoked to explain the dearth of gas below about 1 keV 
(Peterson et al. 2003; Tamura et al. 2001) in the ICM relies on the heating 
by a central AGN. In principle, radio galaxies are energetic enough to halt the cooling in the 
centres of galaxy clusters and explain the high-mass truncation of the galaxy luminosity function 
(see e.g. Best et al. 2006). Radio-loud active galactic nuclei drive strong outflows in the form of 
jets that inflate bubbles or lobes. These lobes are filled with hot plasma, and can heat the cluster 
gas in a number of ways (e.g. Br\"uggen, Ruszkowski \& Begelman 2005; Ruszkowski et al. 2004; 
Br\"uggen \& Kaiser 2002; Churazov et al. 2001). 

Direct evidence for AGN heating is growing. Images taken by CHANDRA and XMM-Newton have shown that AGNs 
in the centres of groups and clusters inflate bubbles in the surrounding X-ray emitting gas. 
Observations of the Perseus Cluster (Fabian et al. 2003; 2006) and of the Virgo Cluster 
(Forman et al. 2005, Simionescu et al. 2006) reveal sound 
waves and weak shocks that extend out to several tens of Kpc. Similar results are reported 
from investigations of Hydra A (Nulsen et al. 2005), MS0735.6+7421 (McNamara et 
al. 2004) and Abel 478 (Sanderson et al. 2004).
From a more statistical point of view, Croston et al. (2005) considered a well-defined 
and homogeneous sample of ROSAT groups and report evidence for the gas properties 
of groups containing radio galaxies to differ from those associated to radio-quiet ones, 
in the sense that radio-loud groups are likely to be hotter at a given X-ray luminosity 
than radio-quiet groups. They attribute this to the effect of heating induced 
by radio galaxies.  Dunn \& Fabian (2006) analysed a more heterogeneous sample of 30 
clusters that show clear signatures of radio activity either in the form 
of 'bubbles' or in the form of a central radio core and study some 
of the properties of AGN heating.

%%%%%%%put back in %%%%%%%%%
%In fact, radio sources in cooling clusters are different from the general population of radio sources in 
%galaxy clusters. The study of Owen \& Ledlow (1997) presents radio and optical data of about 250 radio 
%sources in nearby Abell clusters and reveals that the traditional FR I/FR II classification is not 
%suitable for these sources. Most radio galaxies in clusters are type I sources in terms of their radio 
%power, but they show a great variety of morphologies and dynamics. The central galaxy in almost every 
%strong cooling core contains an active nucleus and a currently active, jet-driven radio galaxy.

More recently, Best et al. (2007) investigated the radio properties of galaxies 
located in the centres of 625 nearby groups selected from the Sloan Digital Sky Survey Data Release 
4 (DR4; Strauss et al. 2002) and argue that AGN heating from the cluster central galaxy 
probably overcompensates the radiative cooling losses in groups of galaxies, therefore 
accounting for the observed entropy floor in these systems. However, Best et al. (2007) did not 
study the thermal properties of the ICM using X-ray data. Lastly, Jetha et al. (2006) examined the 
central regions of 15 galaxy groups with CHANDRA and find that repeated outbursts have a 
long-term cumulative effect on the entropy profiles in their sample.

By following Croston et al. (2005) and Best et al. (2007), the present work 
assesses the importance of (central) radio emission on the thermal properties of the intracluster 
medium in groups and clusters of galaxies from a statistical point of view.
Our approach provides an improvement with respect to previous works as it 
addresses the issue of small statistics (present in Croston et al. 2005), 
lack of information on the thermal properties of the ICM (present in Best et al. 2007) and it 
also extends the investigation of the AGN-cluster properties up to the highest cluster 
masses ($\sim 10^{16}M_\odot$), while both the Best et al. (2007) and Croston et al. (2005) 
analyses only go up to rich groups/poor clusters scales.
Finally, we distinguish radio sources in cluster centres on the basis of 
their radio morphology. As it will be discussed throughout the paper, such a distinction 
has an important impact on the analysis of the general properties of the ICM.

Here, we have considered a statistically significant sample of 145 X-ray selected 
clusters brighter than $3\times 10^{-12}{\rm erg\: s^{-1}cm^{-2}}$, 
taken from the REFLEX and NORAS surveys and endowed with central 
radio emission above the 3 mJy flux limit.
The above limits ensure $\simgt 80 \%$ completeness both in the 
X-ray and radio selections. We investigate the thermal properties of our cluster sample 
and compare them to pre-existing results from the literature. 
We have also made use of detailed Monte Carlo simulations to study 
the behaviour of radio sources surrounded by a dense medium such as that 
of rich clusters.

The layout of the paper is as follows: \S 2 presents the parent catalogues 
from which our sample was drawn, while \S 3 discusses the procedure to assign 
a radio source to cluster centres. \S 4 describes 
the final catalogue, with particular attention to the issue of extended vs 
unresolved radio structures. \S 5 illustrates our results on the fraction 
of X-ray selected clusters which host a central radio source (\S 5.1), 
on the X-ray and radio luminosity relations with cluster temperature 
(\S 5.2), on the X-ray, radio and mechanical luminosity relations with 
the clusters velocity dispersion (\S 5.3) and on the mass deposition rate 
$\frac{dM}{dT}$ (\S 5.4). \S 6 presents the results from Monte Carlo simulations 
on the X-ray luminosity distribution of clusters inhabited by a central radio 
source and 
on the radio luminosity distribution of radio sources surrounded by a dense 
medium. \S 7 summarizes our results and discusses the complex interplay between 
radio source activity and thermal properties of the clusters 
which host them.

Throughout this work we adopt a flat cosmology with a matter
density $\Omega_0=0.3$ and a vacuum energy density
$\Omega_\Lambda=0.7$, a present-day value of the Hubble parameter
in units of $100$ km/s/Mpc,  $h_0=0.7$.

\section{The parent catalogues}
\subsection{NORAS and REFLEX}
The Rosat-Eso Flux Limited X-ray galaxy cluster survey (REFLEX; B\"ohringer et 
al. 2004) provides information on the X-ray properties, fluxes, 
redshifts and some 
identification details for a sample of 447, 
$z\simlt 0.45$, galaxy clusters brighter than an X-ray flux of 
$\sim 2\times 10^{-12} {\rm erg\: s^{-1}cm^{-2}}$ (0.1 to 2.4 KeV). 
It covers 4.24 sr (corresponding to $\sim$ 34\% of the entire sky) 
by observing 
at almost all Galactic latitudes $|b|\ge 20^\circ$ and declinations 
$\delta\le 2.5^\circ$. The catalogue completeness is estimated to be 
$\ge 90$ per cent for fluxes brighter than $3\times 10^{-12} 
{\rm erg\: s^{-1}cm^{-2}}$. The number of sources fulfilling this flux 
requirement is 425.

The Northern ROSAT All-Sky galaxy survey (NORAS; B\"ohringer et al. 2000) 
contains 484 clusters (including the supplements to the original 
survey catalogue) with measured redshifts up to $z\sim 0.45$. This number  
decreases to 245 if one considers the flux limit for  
REFLEX completeness. The survey probes 
Galactic latitudes $|b|\ge 20^\circ$ and declinations $\delta\ge 0^\circ$, and 
is estimated to be $\sim 82$\% complete with respect to the REFLEX survey 
at the same $3\times 10^{-12}{\rm erg\: s^{-1}cm^{-2}}$ flux levels. 

X-ray luminosities in the [0.1-2.4] KeV band are provided for both the REFLEX 
and NORAS clusters. However, while in the case of the REFLEX survey these 
are given in the concordance $\Lambda$CDM cosmology, the NORAS 
catalogue still refers to an EdS model with $h_0=0.5$. In order to 
correct for this problem, 
we converted the NORAS X-ray luminosities from the original EdS 
cosmology to the one adopted in this work by simply writing 
$L^{\Lambda{\rm CDM}}_X= L_X^{\rm EdS}\cdot (x_{\Lambda{\rm CDM}}/x_{\rm EdS})
^2$, where $x$ is the comoving coordinate evaluated at the specific redshift 
of each object. 

For the conversion from X-ray luminosities in the relevant [0.1-2.4] KeV band 
to cluster masses we followed the approach presented by 
Borgani et al. (2001). As a first step, we computed bolometric and 
K-corrections to the [0.1-2.4] KeV-observed band by using the relation 
provided by B\"ohringer et al. (2004).
%a Raymond-Smith 
%(1977) 
%model with Z=0.3 for the mean ICM metallicity. 
Then, for the relation between 
temperature and bolometric luminosity, we considered the phenomenological 
expression:
\begin{equation}
L_{\rm bol}=L_6\left(\frac{T_X}{6{\rm KeV}}\right)^\alpha(1+z)^A
10^{44}h_0^{-2}{\rm erg\; s^{-1}},
\end{equation}
where we took $L_6=3$, $\alpha=3$ (e.g. White, Jones \& Forman 1997; 
Wu, Xue \& Fang 1999) and $A=0$.
Finally, in order to convert temperature into mass, we assumed the 
conditions of virialization, hydrostatic equilibrium and isothermal gas 
distribution to hold by considering the relation:
\begin{equation}
kT=1.38 \beta ^{-1} \left(\frac{M_{\rm vir}}{10^{15}h_0^{-1}}\right)^{2/3}
[\Omega_0\Delta_{\rm vir}(z)]^{1/3} (1+z)\;{\rm KeV},
\end{equation}
(e.g. Eke et al. 1998), where we take 76\% of gas to be hydrogen, 
$M_{\rm vir}$ 
is the cluster virial mass expressed in solar masses, $\beta$ is 
the ratio between the kinetic energy of dark matter and the gas thermal 
energy, and $\Delta_{\rm vir}(z)$ is the ratio between the average density 
$\rho$ within the virial radius and the mean cosmic density at redshift $z$. 
In our case we assume $\beta=1.15$, as found by the Santa Barbara cluster 
comparison project (Frenk et al. 1999). 
Once we have computed the virial masses, we obtain the virial radii 
via $r_{\rm vir}=(3\,M_{\rm vir}
/4\pi\rho)^{1/3}$. Given the uncertainties associated with the parameters 
used in equations (1) and (2) and the errors related to the determination of $L_X$ 
in both the REFLEX and NORAS surveys, 
we note that the values quoted throughout this work 
for both $M_{\rm vir}$ and $r_{\rm vir}$ have to be considered estimates.

\subsection{NVSS}
The NRAO VLA Sky Survey (Condon et al. 1998) is a radio survey that 
has observed the entire sky north of -40 deg declination at 1.4 GHz.
The NVSS was observed with the array in the D configuration (DnC for most 
of the southern sky), which provides an angular resolution of 45 arcsec. 
More than 1.8 million sources were observed down to a flux limit 
$S_{1.4 \rm GHz}=2.5$~mJy, and the survey is shown to be $\sim 80$\% complete at 
fluxes brighter than $3$~mJy. The rms uncertainties in right ascension 
and declination vary from $<$1 arcsec for relatively strong 
($S_{1.4 \rm GHz}\simgt 15$~mJy) point sources to $\sim7$ arcsec for 
the faintest detectable objects.

\section{matching procedure}

As explained in the introduction, the scope of the present work is two-fold: 
first, it aims at investigating the properties of those radio sources 
that reside in the proximity of cluster centres. Second, 
it studies the impact and possible effects of radio sources on the surrounding 
ICM. Note that for our purposes cluster centres are assumed to coincide 
with the centres of X-ray emission, so that in most (but not all) cases 
the adopted centre is coincident with the position of the central cluster 
galaxy (hereafter CG).

In the region where X-ray and 
radio surveys overlap ($\delta\ge -40^\circ$), we compiled a catalogue of X-ray selected clusters 
with enhanced central radio emission by performing a 
cross-correlation analysis between objects brighter than $3\times 10^{-12} 
{\rm erg\: s^{-1}cm^{-2}}$ in the NORAS and REFLEX samples and radio sources 
from the NVSS dataset with 1.4~GHz fluxes above 3~mJy. These flux limits 
were required to ensure a high degree of completeness both in the primary 
X-ray selection and in the search for radio counterparts (cfr. Sections 2.1 
and 2.2). There are 425 clusters from the REFLEX survey fulfilling the 
above condition on X-ray fluxes. This number falls to 305 when we 
consider only those sources that lie at $\delta\ge -40^\circ$, necessary for the 
overlapping of the radio and X-ray surveys.
The total number of clusters used for our analysis is then 550, where 245 
come from the NORAS survey. 

  A suitable matching radius should maximize the number of real associations and
minimize spurious matches. Here we have to consider three factors:  
{\it i)} the wide range of redshifts spanned by clusters in the REFLEX and NORAS surveys 
(from  $z\sim 5\cdot 10^{-3}$ to $z\sim 0.3$), {\it ii)} the fact that 
X-ray and radio emission in the proximity of the cluster centre can be 
displaced from one another and {\it iii)} that NORAS and REFLEX are flux-limited 
surveys probing on average more luminous and more massive clusters at increasing 
redshifts. All these factors converge at indicating that we cannot use a fixed matching 
angular radius. In fact, an angular radius 
suitable for sources around $z\sim 0.2$ would be too large for the 
more local objects and as a result would increase (even by a large factor for the 
very local clusters) the chances of random associations. On the other hand, 
an angular radius suitable for local sources would most likely miss real 
radio-to-X associations at higher redshifts.   

Consequently, we have chosen to adopt a varying 
angular radius in our cross-correlation analysis. 
Since on average clusters of higher mass are found to host more massive 
and therefore larger CGs (e.g von der Linden et al. 2007) -- 
so that the possibility for a greater displacement 
between X-ray and radio emission originating from within the eventual CG is 
higher -- we have adopted a matching radius of the form 
$\theta_{\rm match}=Q\;\theta_{\rm vir}$, 
where $\theta_{\rm vir}$ is the angular extension of the cluster given 
by $\theta_{\rm vir}=r_{\rm vir}/x$, with $r_{\rm vir}$ virial radius 
and $x$ comoving distance.
 
Particular attention was paid to the determination of the value for the
quantity $Q$ used in $\theta_{\rm match}$. We find
that $Q=0.015$ (i.e. $\theta_{\rm match}$ 1.5\%
of the projected virial radius) produces a large enough sample of X-ray
clusters with a central radio counterpart while minimizing contamination
from interlopers. These interlopers can reside both inside
(i.e. radio emitters other than the central source) and
outside (i.e. spurious matches originating from projection effects)
the clusters in our sample.
 
By applying the above matching criteria to the REFLEX and NORAS samples, 
we find that 81 REFLEX clusters out of 305 -- 
corresponding to 26.5\% of the sample -- have a 
radio counterpart that is offset from the centre of X-ray emission by less 
than 1.5\% of the virial radius, while in the case of NORAS we have 67 
radio matches out of 245 objects (i.e. 27.3\% of the original dataset) 
within the same angular distance. 
It is reassuring that the percentage of matches is approximately the same 
in the two cluster surveys. These figures are also in agreement with 
the results from Best et al. (2007) for the occurrence of central radio counterparts 
in their SDSS cluster sample (see their Figure 2). Croston et al. (2005) and Dunn 
\& Fabian (2006) find somehow higher values (on the order of 50\%), but the selection 
criteria for these two samples are different from ours, especially since both groups 
consider systems that are on average at much lower redshifts than our sample.

We note that three sources in REFLEX 
(namely RXCJ0338.4-3526, RXCJ1501.1+0141, RXCJ2347.7-2808) and two in 
NORAS (RXCJ1229.7+0759 and RXCJ1242.8+0241) allowed for more than one 
radio counterpart within the adopted matching radius. Visual inspection of 
radio maps has shown that in all five cases the central 
radio sources had extended/multiple structures. The radio flux 
associated with each of these objects was taken to be the sum of their 
sub-components. We will get back to this point in Section 4.1.

\begin{figure}
%\vspace{8cm}  % amount of vertical space needed
\includegraphics[width=8.0cm]{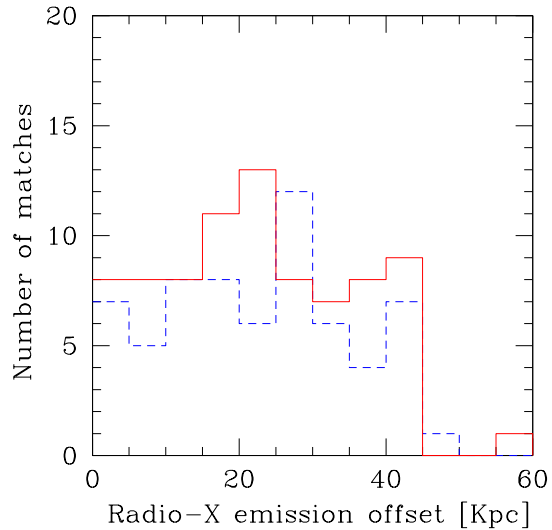} 
%hoffset=-10 voffset=-80 vscale=44hscale=44 angle=0}
%\special{psfile=hist_radio_X_emission.ps hoffset=-10 voffset=-80 vscale=44
%hscale=44 angle=0} 
\caption{Distribution of residuals between radio and X-ray 
positions for objects in the REFLEX (solid line) and NORAS (dashed lines) 
catalogues.
\label{fig:residuals}}
\end{figure}

Chances of contamination from spurious sources in the joint radio--X-ray 
catalogue as obtained above
have been estimated by simply vertically shifting all radio positions by 
1 degree and by re-running the matching procedure. By doing so, we 
find that in the case of both REFLEX and NORAS 
the probability that a radio source falls by chance within the search radius 
from the centre of X-ray emission is about 5.6\%. 
The adoption of a smaller matching radius sensibly reduces the number of 
radio sources found in the proximity of the centre of X-ray emission, 
while leaving the chances for spurious contamination almost 
unaltered. For instance, if instead of taking $Q=0.015$ we consider $Q=0.01$,  
in the case of REFLEX we only obtain 56 matches, while the estimated 
percentage of random coincidences remains almost the same ($\sim 4.5$\%). 
On the other hand, increasing the allowed matching radius by even 
a small amount greatly increases the number of possible spurious 
counterparts: for $Q=0.02$ this figure already gets as high as 13\%.

The distribution of residuals between radio and X-ray positions as a 
function of the (comoving) distance from the centre of the cluster is 
shown in Figure 1.
The solid line indicates the case for REFLEX clusters, while the dashed line 
refers to the NORAS sample. All radio positions differ from the 
X-ray ones by less than $\sim 50$~Kpc. If a central cluster galaxy is present, 
this implies -- as expected -- 
that radio emission originates from it. 

The number of clusters brighter than $3\cdot 10^{-12}$ erg s$^{-1}$cm$^{-2}$ 
endowed with a central radio counterpart above the 3~mJy level as derived 
from our analysis is 148. Three clusters, however, have multiple 
identifications in the NORAS and REFLEX catalogues (i.e. appear in both surveys). 
By removing these sources 
from the radio-NORAS sample, we end up with 145 clusters with a unique central radio 
counterpart. This is the sample that will be used throughout this work. 

\section{Description of the sample}
Relevant properties for the radio sources found in the proximity of 
NORAS and REFLEX cluster centres by following the procedure described in 
Section 3 are summarized in Table 2 which reports in the various columns:\\
(1) Name of the parent cluster.\\
(2)-(3) RA(2000) and Dec(2000) of the radio source.\\
(4) Radio flux expressed in mJy ($F_{1.4 \rm GHz}$).\\
(5) Log of the radio luminosity expressed in W/Hz/sr (${\rm log_{10}} L_R$).\\
(6) Redshift of the cluster ($z$).\\
(7) Distance between cluster centre and radio counterpart (Kpc).\\ 
(8) Notes on the radio appearance, i.e. whether point-like (blank space) 
or extended/sub-structured source (denoted with an 'S'). 

Radio luminosities have been calculated according to the relation:
\begin{eqnarray}
L_{\rm 1.4 GHz}=F_{\rm 1.4 GHz} D^2 (1+z)^{3+\alpha},
\label{eq:rlum}
\end{eqnarray}
where $D$ is the angular diameter distance and $\alpha$ is the spectral index 
for radio emission [$F(\nu)\propto \nu^{-\alpha}$]. Since we do not have measurements 
for $\alpha$, we assumed $\alpha=0.75$ which is the typical 
value found for early-type galaxies and steep spectrum sources 
such as jets and lobes. Note that, since the maximum redshift 
probed by both cluster surveys is always $\simlt 0.3$, estimates 
from equation (3) do not depend on the precise value of the spectral index. 

\subsection{Radio sources with extended emission}

The issue of extended/sub-structured sources (point (8) in Table 2) deserves 
a digression. As mentioned in Section 3, five clusters are associated with radio 
sources that have multiple components within the 
chosen search radius. A visual investigation of the radio images taken 
from the NVSS survey for all the 145 radio objects associated with 
NORAS and REFLEX clusters shows that about 42 per cent of them differ from 
a point-like structure. (Note that since the NVSS resolution is $45^{\prime\prime}$,  
multiple structures closer than $\sim 9$~Kpc at $z=10^{-2}$ up to $\sim 250$~Kpc
at the highest redshifts probed by the REFLEX and NORAS surveys 
will be unresolved in the NVSS radio maps. We will get back to this point 
in Sections 5.1 and 5.2). 
More precisely, this happens for 61 sources, out of 
which 33 are found in REFLEX clusters and 28 in NORAS clusters. 
The percentage quoted above is much higher than what is found for 'field' radio 
galaxies within the same redshift range and at similar radio flux levels 
($\sim 5\%$ cfr. Magliocchetti et al. 2001). We will get back to this point in Sections 6 and 7.

Having labelled the radio sources with an 'S', we do not make any 
further distinction on whether the radio emission presents extended blobs, 
it appears as a close or distant double-lobed object or features a triple/more 
complex structure. For the purposes of the present work, the only division   
we are interested in is whether the radio source exhibits signs of 
extended emission or not.
  
Generally, it was relatively easy to recognize different sub-structures that belong 
to the same radio source. Nevertheless, there are a number of cases that remain uncertain 
and that have been labelled with an 'S?' in Table 2. To play safe, these sources 
have not been included in the "sub-structure" sample and in the following analyses 
they will be treated as point-like objects.   

Radio fluxes for extended radio sources in Table 2 have been obtained by adding up 
all the fluxes of their different components. Radio luminosities have then 
been computed by means of equation (\ref{eq:rlum}) for the resulting total fluxes. 

\section{Observational properties of the sample}

In this section, we will show the most interesting properties and relations of our sample. 
In addition to the data described in Section 4, some extra information such as temperatures 
and velocity dispersions has been gathered from the BAX and VIZIER databases.  All the best-fit values 
for the relations presented in the following pages are summarized in Table 1, while we defer a detailed 
interpretation of the results to Sections 6 and 7.

\begin{table*}
\begin{center}
\caption{Best-fit values for the trends presented in Section 5. All relations are of the 
kind ${\rm log}_{10} y=b {\rm log}_{10}x +a$. X-ray luminosities $L_X$ are 
expressed in [10$^{44}$ erg/s], radio luminosities $L_R$ in [W/Hz/sr], 
temperatures $kT$ in [KeV], velocity dispersions $\sigma$ in [Km/s] and mass 
deposition rates $\frac{dM}{dt}$ in [M$_\odot$/yr]. In brackets are 
the number of sources used to evaluate the various fits.}
\begin{tabular}{llll}
\hline
\hline

              &   Point-like sources    & Extended Sources  & Whole sample \\
\hline
\hline
$L_X$ vs $kT$ (Fig. 7a) 
& $b=2.35^{+0.08}_{-0.07}$; $a=-1.02^{+0.04}_{-0.04}$ (33) & 
$b=4.0^{+0.2}_{-0.2}$; $a=-2.17^{+0.10}_{-0.09}$ (31) 
& $b=2.8^{+0.1}_{-0.1}$; $a=-1.29^{+0.06}_{-0.05}$ (64)\\
\hline
$L_R$ vs $kT$ (Fig. 7b)& 
$b=5.6^{+0.2}_{-0.2}$; $a=20.40^{+0.1}_{-0.09}$ (33) &  
$b=7.2^{+0.4}_{-0.3}$; $a=19.5^{+0.1}_{-0.2}$ (31) & 
$b=6.3^{+0.3}_{-0.2}$; $a=20.1^{+0.1}_{-0.1}$ (64)\\
\hline
$L_X$ vs $\sigma$ (Fig.8a) & 
$b=6.2^{+0.4}_{-0.3}$; $a=-17.2^{+1.0}_{-2.4} $(19)& 
$b=5.5^{+1.0}_{-0.8}$;  $a=-15.7^{+2.7}_{-3.1}$ (20)& 
$b=3.8^{+0.4}_{-0.4}$; $a=-10.6_{-1.0}^{+0.9}$ (39)\\
\hline
$\sigma$ vs $kT$ (Fig. 8b) &$b=0.44^{+0.04}_{-0.06}$; $a=2.55^{+0.03}_{-0.01}$ (16)& 
$b=0.53^{+0.07}_{-0.11}$; $a=2.54^{+0.06}_{-0.02}$ (18)&
$b=0.49^{+0.03}_{-0.09}$; $a=2.55^{+0.04}_{-0.01}$ (34)\\
\hline
$L_{\rm mech}$ vs  $\sigma$ (Fig. 9) & $b=-6.8\pm 0.3$; $a=17.6\pm 0.9$ (19) 
&$b=-6.9 \pm 0.5$; $a=18.6\pm 1.5$ (20)& $b=-7.1\pm 0.3$; $a=18.3\pm 0.9$ (39)\\
%\hline
%$\frac{dM}{dt}$ vs $L_X$ (Fig. 9a) & -- & -- & $b=1.09\pm 0.07$; $a=?$   (15) \\
\hline
$\frac{dM}{dt}$ vs $L_R$ (Fig. 10) & -- & -- & $b=0.25\pm 0.02$ (15)\\
\hline
\end{tabular}
\end{center}
\end{table*}

%%%%%%%%%%%%%%%%%%%%%%%%%%%%%%%%%%%%%%%%%%%%%%%%%%%%%%%%%%%%%%%%%%%%%%%%%%%%%%%%%%%%

\subsection{Fraction of radio detections}

\begin{figure}
%\vspace{8cm}  % amount of vertical space needed
\includegraphics[width=8.0cm]{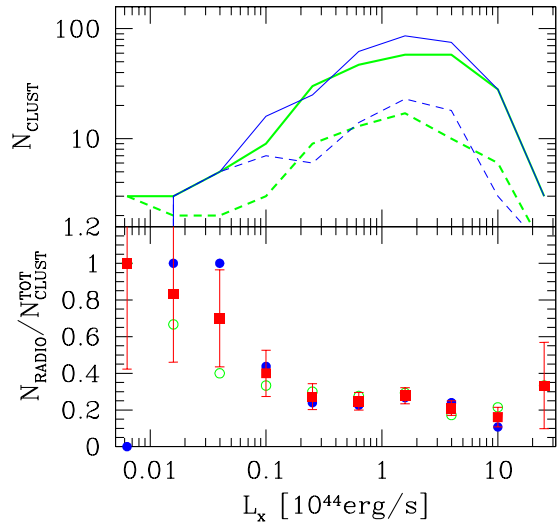}
%\special{psfile=fraclumradio_theta0.015.ps hoffset=-10 voffset=-80 vscale=44
%hscale=44 angle=0} 
\caption{Top panel: number of clusters per X-ray luminosity 
interval. The thick (green) lines illustrate the case for NORAS, while the thin 
(blue) ones are for the REFLEX sample. Solid lines show the entire $F_X\ge 
3\times 10^{-12} {\rm erg\: s^{-1}cm^{-2}}$ population, while dashed ones 
only represent those objects which exhibit relevant ($F_{1.4 \rm GHz}\ge 3$~
mJy) central radio emission. Lower panel: ratio between number of clusters 
with a central radio counterpart and the whole $F_X\ge 
3\times 10^{-12} {\rm erg\: s^{-1}cm^{-2}}$ cluster population. Open (green) 
circles are for NORAS objects, filled (blue) dots for REFLEX, while (red) squares 
and associated errorbars represent the combined REFLEX+NORAS dataset.     
\label{fig:frac_lum}}
\end{figure}

\begin{figure}
%\vspace{8cm}  % amount of vertical space needed
\includegraphics[width=8.0cm]{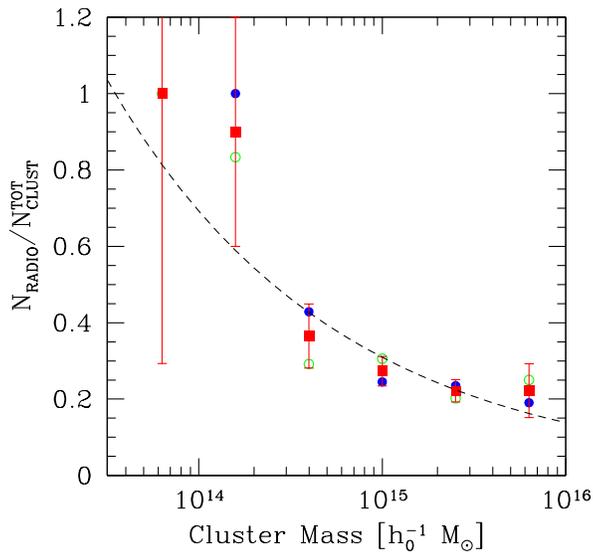}
%\special{psfile=fracmassradio_theta0.015.ps hoffset=-10 voffset=-80 vscale=44
%hscale=44 angle=0} 
\caption{Ratio between number of clusters 
with a central radio counterpart brighter than 3~mJy in the NVSS maps  
and the whole $F_X\ge 
3\times 10^{-12} {\rm erg\: s^{-1}cm^{-2}}$ cluster population as a function 
of virial cluster mass. Open (green) circles 
are for NORAS objects, filled (blue) dots for REFLEX, while (red) squares and 
associated errorbars represent the combined REFLEX+NORAS dataset. The dashed 
line indicates the best fit to the data 
(see text for details). 
\label{fig:frac_mass}}
\end{figure}

\begin{figure}
%\vspace{8cm}  % amount of vertical space needed
\includegraphics[width=8.0cm]{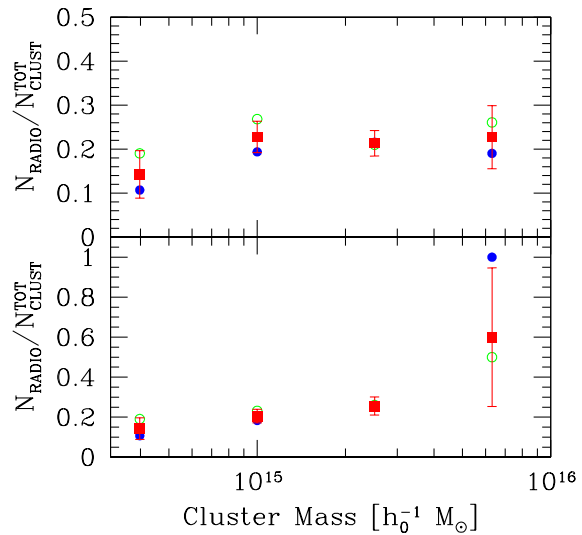}
%\special{psfile=fracmassradio_theta0.015_Plim.ps hoffset=-10 voffset=-80 
%vscale=44 hscale=44 angle=0} 
\caption{Top panel: fraction of X-ray selected clusters with fluxes 
$F_X\ge 3\times 10^{-12} {\rm erg\: s^{-1}cm^{-2}}$ which present 
central radio emission brighter than a luminosity $L_R=10^{21.9}$~
W/Hz/sr as a function of virial cluster mass.  Open (green) circles 
are for NORAS objects, filled (blue) dots for REFLEX, while (red) squares 
and associated errorbars represent the combined REFLEX+NORAS dataset. 
Bottom panel: same as above but for the volume-limited sample 
$L_R\ge 10^{22}$~W/Hz/sr, $z\le 0.14$ (cfr. Fig. \ref{fig:zlimits}b). 
\label{fig:frac_mass_Plim}}
\end{figure}

As an interesting first exercise, we can estimate the number of X-ray clusters brighter than 
3 erg s$^{-1}$ cm$^{-2}$ that host a central radio source above $F_{\rm 1.4 GHz}=3$ 
mJy. Figure \ref{fig:frac_lum} shows, as a function of X-ray 
luminosity, the total number of X-ray clusters used in this work 
(solid lines in the top panel of Figure 
\ref{fig:frac_lum}; the thick line represents the NORAS case while the thin one 
is for REFLEX clusters), the number of X-ray clusters with a central radio 
counterpart (dashed lines in the top panel of Figure \ref{fig:frac_lum}), and the ratio 
between these two quantities (bottom panel of Figure \ref{fig:frac_lum}, where 
squares and associated errorbars indicate the combined REFLEX+NORAS case). 
Note that there is a clear preference for radio sources to appear in poor clusters. 
In fact, despite the relatively poor statistics at low luminosities,  the 
percentage of clusters hosting a central radio-AGN brighter than 3~mJy drops from about 
[50\%-100\%] for $L_X \simlt 10^{43}$~erg/s to $\sim 10$\% at the highest luminosities 
probed by NORAS and REFLEX. 

This trend becomes even more intriguing if one considers the fraction of 
clusters that present central radio emission above 3~mJy as a function of the 
cluster virial mass evaluated as in Section 2.1 (Figure \ref{fig:frac_mass}). 
In this case, there is a trend for central radio detections in clusters to 
decrease with cluster mass; we find that {\it all but one} 
of the 12 $M_{\rm vir}\simlt 10^{14.5} {\rm M_\odot} h_0^{-1}$ 
poor clusters and groups identified in 
NORAS and REFLEX are associated with a central radio emitting source brighter than 3 mJy in 
the NVSS maps, while this figure rapidly decreases to a $\sim 20$ per cent as soon 
as one moves to masses $M_{\rm vir}\simgt 10^{15} {\rm M_\odot} h_0^{-1}$. 
We can quantify this result by writing: ${\rm f_R=N_{RADIO}/N^{TOT}_{CLUSTERS}}
=\left(M_{\rm vir}/M_*\right)
^{-\alpha}$, where $M_*=(3.5\pm 0.9)\cdot 10^{13} {\rm M_\odot} h_0^{-1}$ and 
$\alpha=0.35 \pm 0.08$ (dashed line in Figure \ref{fig:frac_mass}).

Obviously, selection effects bias the results presented in Figures \ref{fig:frac_mass} and 
\ref{fig:frac_lum}. The adopted flux limit for NORAS and REFLEX implies 
that X-ray luminous and therefore more massive clusters are preferentially 
observed at higher redshifts. 
On the other hand, the radio-selection function derived from having used a 3~mJy 
limit for our radio sample (solid line in Figure \ref{fig:zlimits}b) implies that while 
locally one is capable of detecting all radio sources brighter than 
$\sim 10^{19.5}$ W/Hz/sr, at the redshifts probed by massive clusters only radio sources 
brighter than $\sim 10^{22}$ W/Hz/sr can be observed. 
Furthermore, probing the most luminous clusters at high redshifts may 
imply selecting relaxed clusters i.e. those which are cool-core ones.  
If radio-loud AGN are the origin for any feedback, then this may be an extra 
selection effect, in that cool-core clusters are more likely to host powerful 
radio sources.
 
Nevertheless, the observational fact that all but one of the local clusters and groups  
host a central radio-luminous source suggests that -- by going deep enough in radio maps -- 
this might also be the case for the majority of X-ray selected clusters.  This 
conclusion becomes more stringent by the fact that our criterion for associating radio 
sources with cluster centres necessarily misses relatively common objects such as 
triple structures where the central component is too faint in radio to be seen 
(double-lobed sources with a 'radio-quiet' galaxy set in their 
middle).

To obtain an unbiased trend for the recurrence of radio sources in cluster centres, 
we compile a volume-limited sample. This is presented in the 
bottom-panel of Figure \ref{fig:frac_mass_Plim} for the volume-limited sample 
$z\le 0.14$, $L_R\ge 10^{22}$ W/Hz/sr (cfr. Figure \ref{fig:zlimits}b), chosen 
such that the number of sources available for the analysis is maximised. 
In this case, the fraction of X-ray clusters with central radio emission 
is found to be around the 20 per cent level, with possibly a slight preference for radio 
sources to inhabit more massive clusters. We note however that this result only holds for 
relatively large clusters as the radio luminosity cut has excluded all clusters less massive 
than $M_{\rm vir}\sim 10^{14.5} {\rm M_\odot} h_0^{-1}$.

Best et al. (2007) use a sample of 625 groups and clusters of galaxies selected from 
the Sloan Digital Sky Survey and cross-correlated with the NVSS and FIRST (Faint 
Images of the Radio Sky at Twenty centimeters; Becker et al. 1995) radio samples to study 
the properties of radio-loud AGN in the brightest cluster galaxies (coinciding in the 
majority of the cases with our definition of CGs, see von der Linden et al. 2007). 
In their work they show that the fraction of brightest cluster galaxies that are   
radio-loud AGN increases with (stellar) mass as $\propto M_{\rm stellar}^{1.0}$ up to a 
plateau level of about $\sim 20-30$ per cent reached for their highest mass range 
which is centred at around $M_{\rm stellar}\sim 10^{11.5} M_\odot$. \\
Our sample of X-ray selected clusters allows to extend the Best et al. (2007) analysis 
to the higher-mass region probed by rich groups and clusters of galaxies. The trend 
presented in the top panel 
of Figure \ref{fig:frac_mass_Plim} then shows the fraction of clusters with a central radio 
counterpart brighter than $L_R\ge 10^{21.9}$ W/Hz/sr (limit which is comparable to the 
definition of radio loudness given in Best et al. 2005) as a function of 
cluster mass. At all cluster masses, the percentage of clusters hosting a central 
radio-loud source is substantially constant and equal to $\sim 20$ per cent. 
Our results, both for what 
concerns the presence of a plateau in the distribution of objects that are radio-loud AGN 
and for the value of such fraction, are then in agreement with those of Best et al. 
(2007). It is quite intriguing that the same dependence (or rather independence) on mass 
of the occurrence of central radio-loud sources in massive extragalactic objects is found 
independently of whether 
such a mass is that of a 'field' galaxy (see Best et al. 2007), of a cluster central galaxy 
or of the cluster that surrounds it.  

\begin{figure}
%\vspace{8cm}  % amount of vertical space needed
\includegraphics[width=8.0cm]{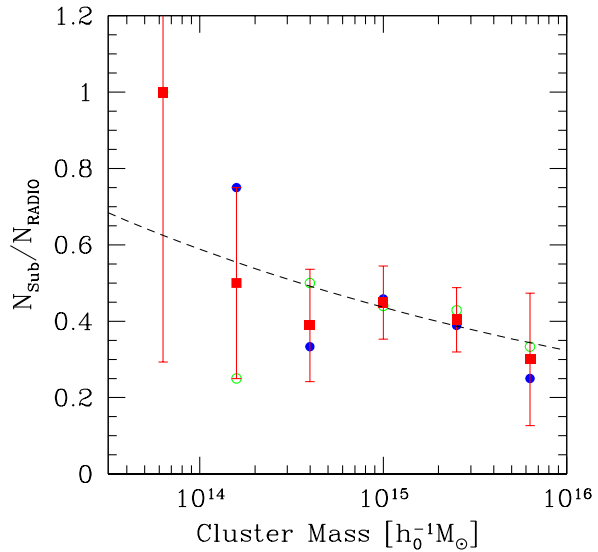}
%\special{psfile=fracmassdouble_theta0.015.ps hoffset=-10 voffset=-80 vscale=44
%hscale=44 angle=0} 
\caption{Fraction of radio-detected 
clusters that present extended/sub-structured radio emission as a function 
of virial cluster mass. Open (green) circles 
are for NORAS objects, filled (blue) dots for REFLEX, while (red) squares and 
associated errorbars represent the combined REFLEX+NORAS dataset. 
The dashed 
line indicates the best fit to the data 
%${\rm f_S=N_{Sub}/N_{RADIO}}
%=\left(M_{\rm vir}/1.6\cdot10^{12}M_\odot\right)^{-0.1}$ 
(see text for details). 
\label{fig:frac_double}}
\end{figure}

\begin{figure}
%\vspace{8cm}  % amount of vertical space needed
\includegraphics[width=8.0cm]{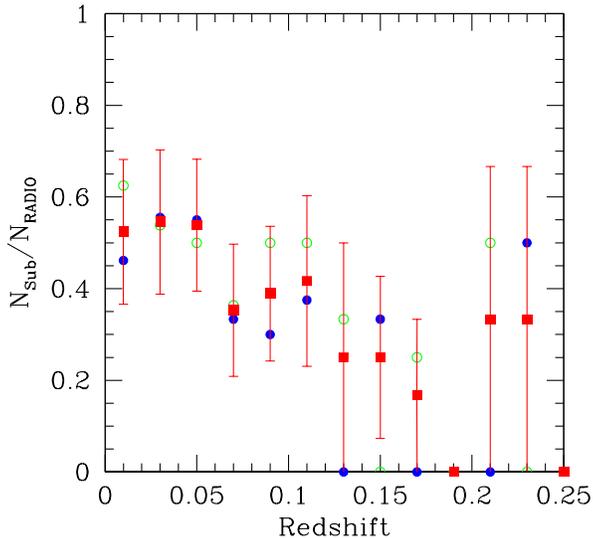}
%\special{psfile=hist_z_doubles.ps hoffset=-10 voffset=-80 vscale=44
%hscale=44 angle=0} 
\caption{Fraction of radio-detected 
clusters that present extended/sub-structured radio emission as a function 
of redshift. Open (green) circles 
are for NORAS objects, filled (blue) dots for REFLEX, while (red) squares and 
associated errorbars represent the combined REFLEX+NORAS dataset. 
\label{fig:frac_double_z}}
\end{figure}

Another interesting point concerns the intermittency of the radio-loud AGN phenomenon.
Best et al. (2005) argue that the observational finding that $\sim 20$-30 
per cent of all the most massive galaxies in their sample present signatures 
of radio activity implies that this activity has to be re-triggered so often that the 
galaxy spends about a quarter of its life-time in an active state. 
This fuel-supply requirement becomes even more demanding if one 
considers that the Best et al. (2005) selection includes radio sources that are 
brighter than those 
considered in this work. As already discussed, Figure \ref{fig:frac_mass} suggests that 
if one goes to low-enough radio luminosities, it is likely that the majority of clusters 
host a central radio source, just as it is found locally for our sample. 
This result would then imply an almost constant re-triggering of the radio source, so to 
allow the hosting galaxy to spend almost all its lifetime in an active state.

As a final point in this section, we have considered those radio sources within 
the cluster centres that present extended and/or sub-structured radio emission and 
investigated whether there was a dependence of their recurrence rate on cluster mass.
The results are presented in Figure \ref{fig:frac_double}. The very marginal trend 
for their fraction to decrease with $M_{\rm vir}$ (${\rm f_S=N_{Sub}/N_{RADIO}}
\simeq\left(M_{\rm vir}/1.6\cdot10^{12}h_0^{-1} M_\odot\right)^{-0.1}$) is most likely due 
to the combined effect of the resolution capabilities of the NVSS survey with 
the already mentioned selection biases in flux-limited surveys which preferentially 
identify massive clusters at high redshifts.
This implies that, at variance with what happens 
more locally, close substructures will not be observed as such at those redshifts, 
resulting in a net loss of sub-structured/extended sources starting from $z\sim 0.05$ and 
particularly beyond $z\sim 0.1$ (cfr. Figure~\ref{fig:frac_double_z}).
Nonetheless, in the local universe 
and/or at low cluster masses there is a tendency for the majority ($> 50$\%) of central 
radio 
sources to possess extended radio emission. Combining this result with that of Figure 
\ref{fig:frac_mass} we can then conclude that, at least locally, the majority of 
clusters (which in this case are rather groups) host an extended radio source 
in their centre.
We will investigate the consequence of this finding in the next sub-sections.

\subsection{Luminosity vs Temperature}

\begin{figure*}
%\vspace{8cm}  % amount of vertical space needed
\includegraphics[width=8.0cm]{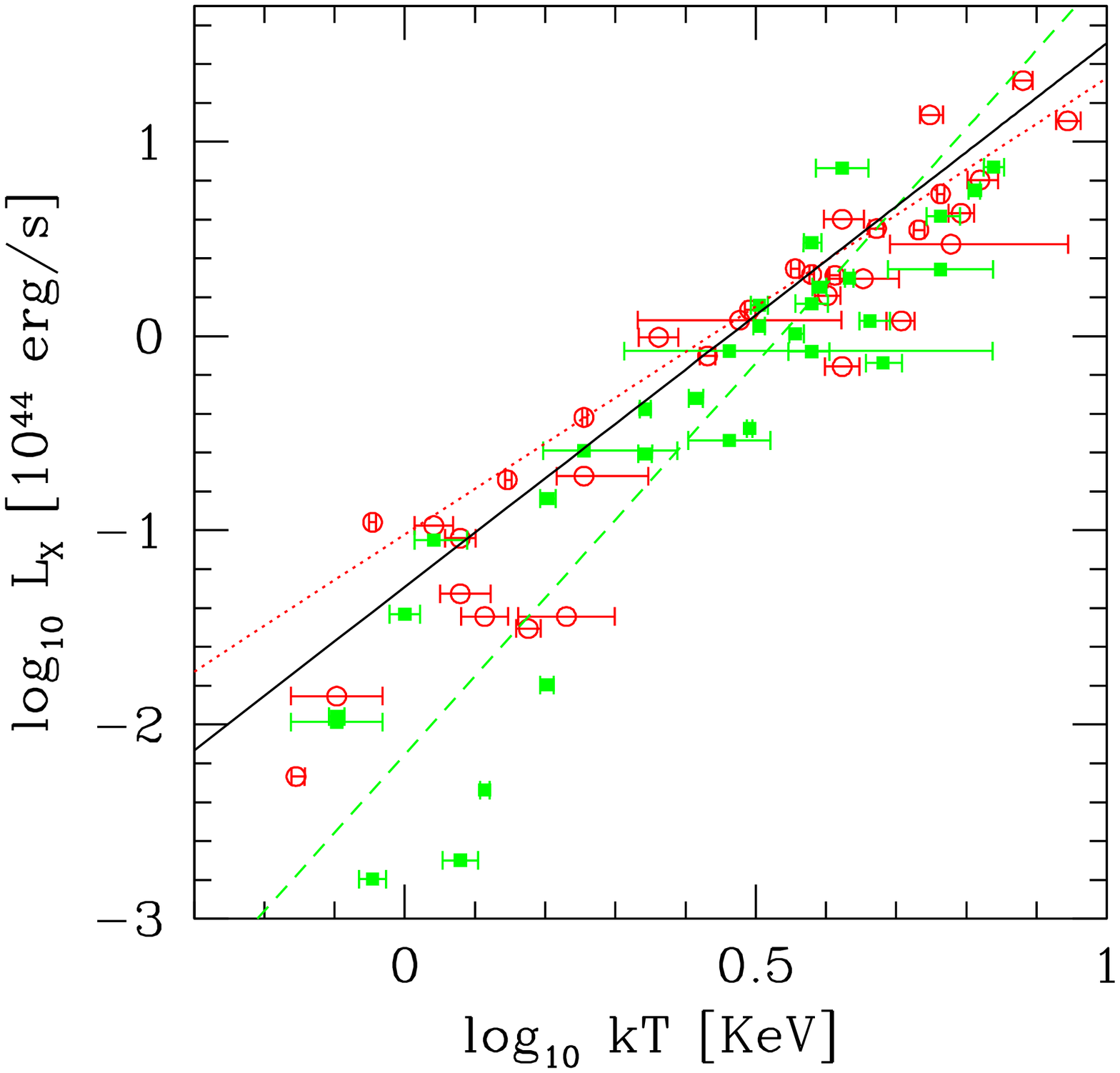}
\includegraphics[width=8.0cm]{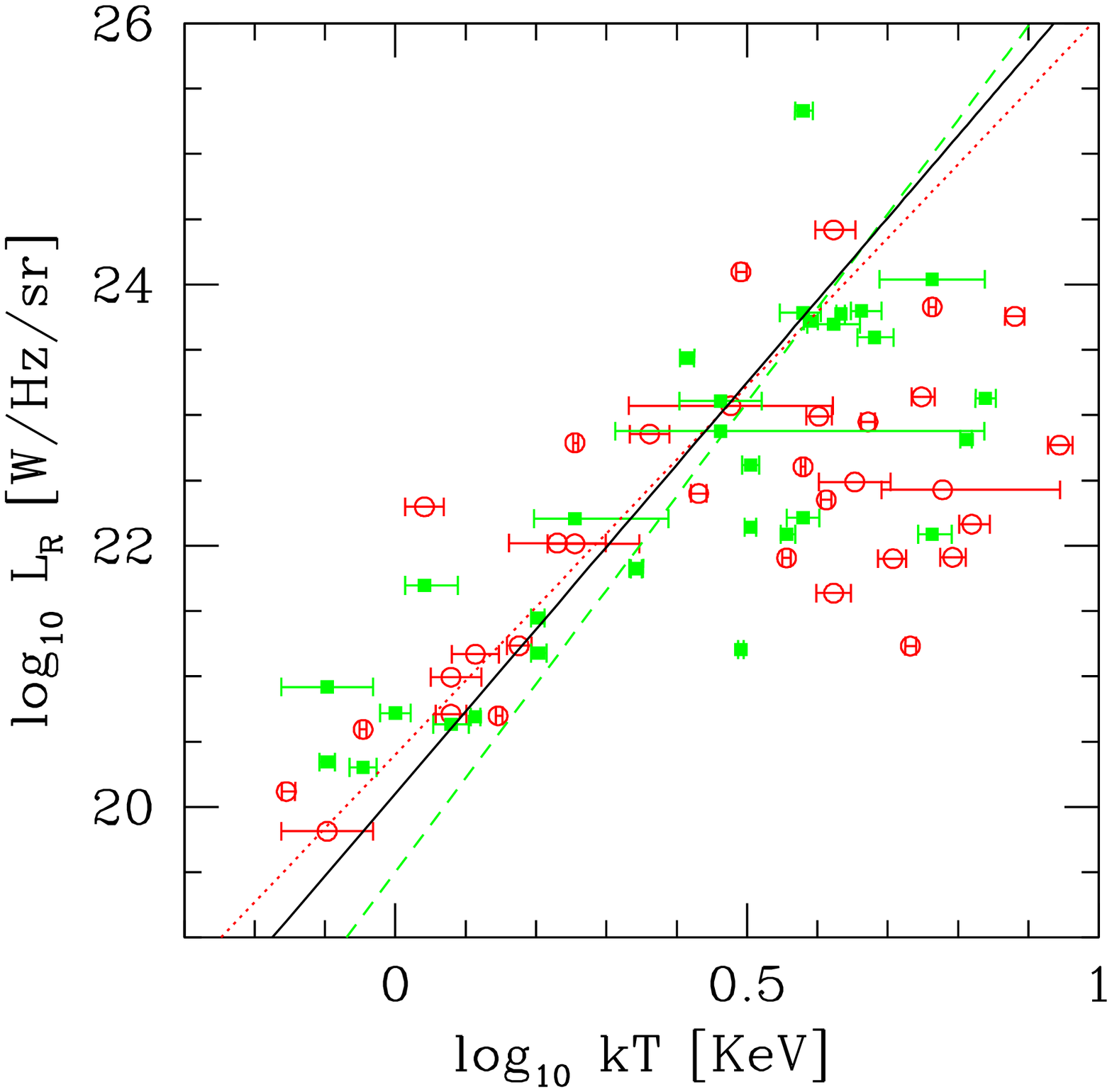}
%\special{psfile=LxvsT_structnostruct.ps hoffset=-10 voffset=-80 vscale=44
%hscale=44 angle=0} 
%\special{psfile=LrvsT_structnostruct.ps hoffset=250 voffset=-80 vscale=44
%hscale=44 angle=0}
\caption{Left-hand panel (a): X-ray luminosity versus cluster temperature for those X-ray 
selected clusters brighter than 3 erg s$^{-1}$ cm$^{-2}$ which exhibit central 
radio emission above 3 mJy. Open (red) circles represent clusters associated to point-like radio 
sources, while solid (green) squares are for those clusters inhabited by radio objects presenting 
extended structures. The dashed (green) line indicates the $L_X-kT$ best fit for the sub-population 
of extended 
radio sources, while the dotted (red) one is for point-like radio objects (see Table 1).  
The solid line is the best fit to the whole radio cluster sample. 
Right-hand panel (b): radio luminosity of the sources associated 
with cluster centres versus cluster temperature. Symbols are as before.
\label{fig:LvsT}}
\end{figure*}

\begin{figure*}
%\vspace{8cm}  % amount of vertical space needed
\includegraphics[width=8.0cm]{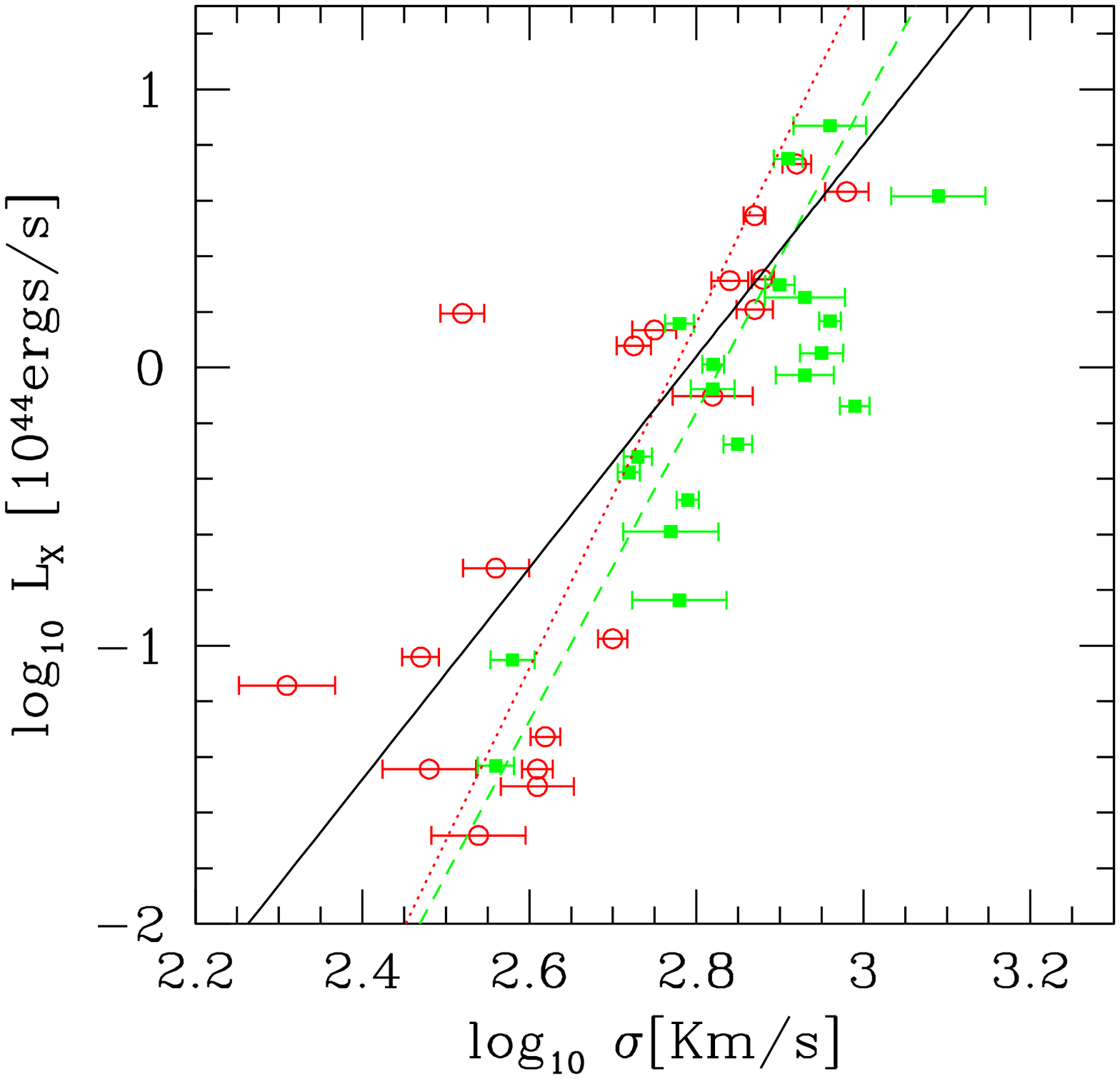}
\includegraphics[width=8.0cm]{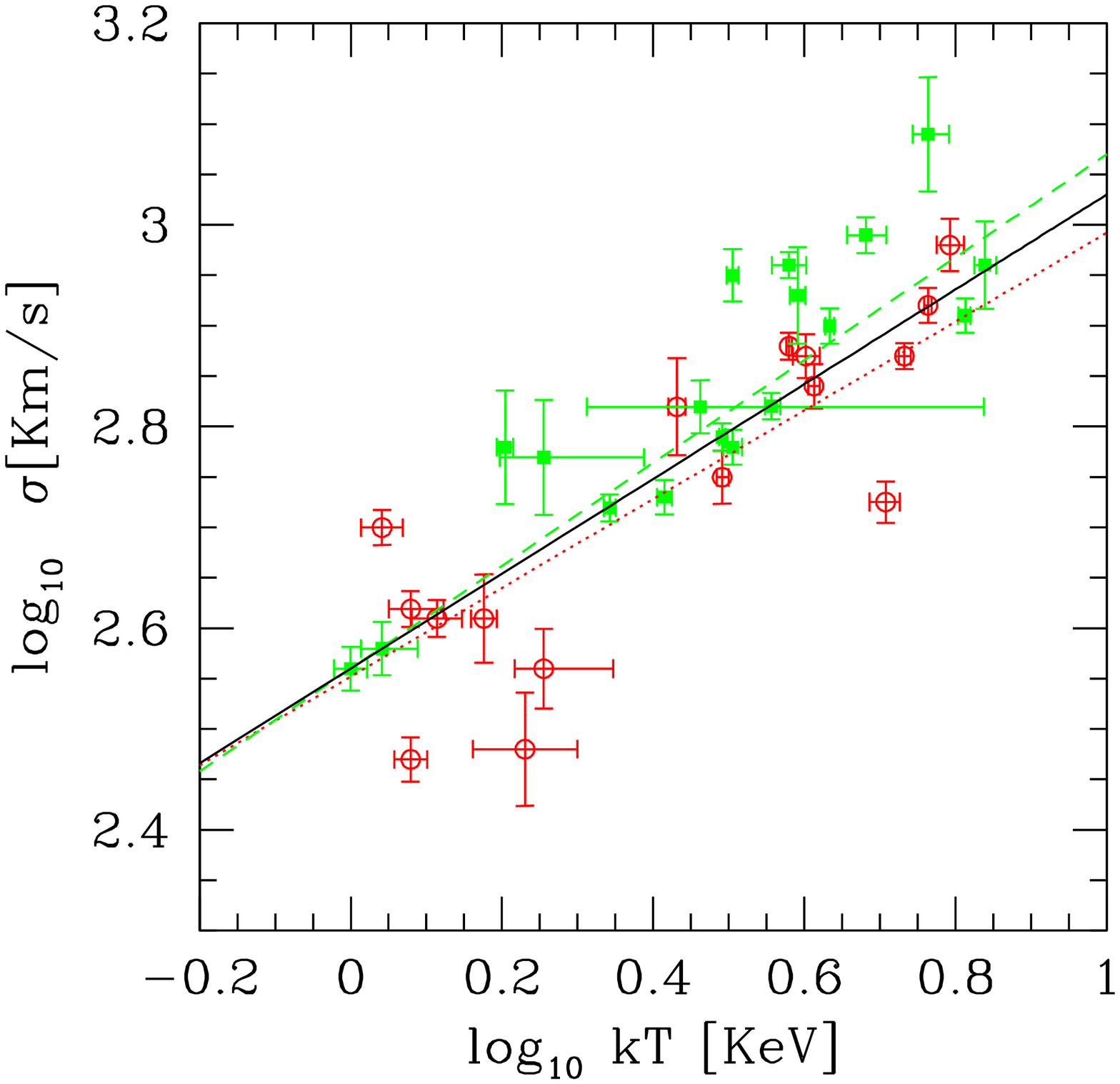}
%\special{psfile=Lxvssigma.ps hoffset=-10 voffset=-80 vscale=44
%hscale=44 angle=0} 
%\special{psfile=tempvssigma.ps hoffset=250 voffset=-80 vscale=44
%hscale=44 angle=0}
\caption{Left-hand panel (a): X-ray luminosity $L_X$ versus velocity dispersion $\sigma$ for 
a subsample of 39 clusters with enhanced central radio emission. Open (red) circles identify 
point-like radio sources while filled (green) squares represent extended ones. The (green) dashed 
and (red) dotted 
lines respectively indicate the $L_X-\sigma$ best fits to the sub-populations of extended 
and point-like radio sources (see Table 1). The solid line is 
the best fit to the whole radio cluster sample. Right-hand panel (b): 
velocity dispersion versus cluster temperature. Symbols are as before. 
\label{fig:Lvssigma}}
\end{figure*}

X-ray scaling relations of galaxy clusters, such as the X-ray
luminosity-temperature relation, play an important role when using clusters of
galaxies to constrain cosmological parameters. When gravitational heating is
the only source of heating, clusters should behave self-similarly and the
X-ray luminosity (due to thermal bremsstrahlung) should scale as $L_X \propto T^2$
(Kaiser 1986).

Most authors (e.g. White, Jones \& Forman 1997; 
Arnaud \& Evrard 1999 and more recently Popesso et al. 2005) report $L_X-kT$ 
correlations with 
logarithmic slopes close to 3 in the cluster regime, though attempts to remove the 
effects of 
central cooling flows (Allen \& Fabian 1998; Markevitch 1998) results in flatter 
relations. 
However, the above result is found to break at the mass/luminosity scales of groups 
of galaxies. Helsdon \& Ponman (2000a,b) report a steeper slope of $4.9\pm 0.8$ for a sample 
of X-ray bright loose groups, and $4.3\pm 0.5$ for a larger sample including both loose 
and compact groups, while Xue \& Wu (2000) found a slope $5.6\pm 1.8$ from data for 38 
groups 
taken from the literature.  More recently, Osmond \& Ponman (2004) -- by analysing 
a sample of 60 galaxy groups within the GEMS project -- have measured an $L_X-kT$ slope 
that is comparable with that found in clusters (but see their 
discussion about possible systematics which could have affected their result).   

The observed departure from self-similarity has been interpreted as excess of 
entropy in low mass clusters due to some form of pre-heating generated by 
sources such as supernovae and/or AGN (see e.g. Tozzi \& Norman 2001). 
Evidence for heating excess in 19 groups containing radio-loud AGN has
recently been reported by
Croston et al. (2005). From the theoretical point of view a number of works
have shown that
radio-loud AGN can provide a sufficiently energetic and widely distributed
heating mechanism
to solve the entropy floor problem (e.g. Br\"uggen \& Kaiser
2001; Br\"uggen \& Kaiser 2002; B\"ohringer et al. 2002; Fabian et al. 2003;
Heinz et al. 2006).

In order to investigate the effects of the presence of a central radio
  component in groups and clusters of galaxies from a statistical point of
  view, we have gathered temperatures for 64 of the radio-detected clusters
  belonging to our sample from the BAX database.  This sample was sub-divided
according to whether the central radio source presented an extended (31
objects) or an unresolved (33 objects) structure.  The $L_X$ vs $kT$ trend for
the two sub-populations is presented in the left-hand panel of Figure
\ref{fig:LvsT}, where open circles represent clusters inhabited by
point-like radio sources, while filled squares are for clusters associated
with central sources with extended radio emission. The difference in behaviour
between these two populations is striking: at a given temperature, the
overwhelming majority of clusters associated with extended radio sources have
a lower X-ray luminosity than those that host
point-like radio objects. Or, from a different point of view, at a given 
luminosity clusters associated with extended radio sources are 
systematically hotter than those that are not.

When we make a $\chi^2$ fit to our entire sample of X-ray selected clusters
that host a central radio source, we find a slope of the $L_X-kT$ relation of
$b=2.8\pm 0.1$. This slope becomes appreciably shallower
($b=2.35^{+0.08}_{-0.07}$) if one only includes unresolved radio sources, and
remarkably steeper ($b=4.0\pm 0.2$) for the sub-population of extended radio
objects.  Intriguingly, X-ray clusters with a point-like central radio
source trace a distribution that is very similar to the $L_X\propto T^2$
expected for self-similar systems. The presence of extended radio sources breaks this
relation causing a much steeper dependence of X-ray luminosity on cluster
temperature. This may point towards a more efficient mode with which extended
sources can permeate the surrounding ICM and transfer heat within the cluster. As
will be discussed in greater detail in Sections 6 and 7, the absolute effect of extended
radio-loud AGN on the ICM decreases with cluster mass.  This is clearly seen in Figure
\ref{fig:LvsT}a. With increasing mass (i.e. increasing $T$), the difference between
point-like and extended radio sources decreases.

The issue of point-like vs extended radio sources has been investigated further by 
considering volume-limited samples of clusters. In fact, given the 45$^{\prime\prime}$ 
resolution of NVSS (cfr Section 4.1 and Figure~\ref{fig:frac_double_z}), the above result on the different 
behaviour of the 
$L_X-kT$ relationship for clusters that host compact or extended radio sources could 
be biased by the presence of radio structures that are extended but 
appear as point-like in the NVSS maps. 
We have then repeated the above analysis by splitting the samples 
into sources with $z\leq 0.05$ and $0.05 < z\leq 0.2$. This means that 
for the closer sub-group of clusters, all radio structures more extended than 
$\sim$46~Kpc will be correctly identified as such in the radio maps. 
However, in the subsample made of more distant objects one will identify as point-like 
all those extended radio sources that have structures smaller than $\sim 190$~Kpc. 
In the former case, the best-fit values for the $L_X$ vs $kT$ relationship are 
$b=4.4\pm0.2$; $a=-2.25\pm 0.08$ for extended sources and 
$b=2.29\pm0.04$; $a=-1.02\pm 0.02$ for point-like structures. 
At higher redshifts one instead gets 
$b=2.9^{+0.8}_{-0.6}$; $a=-1.5^{+0.4}_{-0.6}$  
and $b=2.7\pm 0.3$; $a=-1.3\pm 0.2$. 
One can then see that the results obtained for the more reliable local subsets of clusters  
are in very good agreement with those derived 
previously in this Section (cfr Table 1) regardless 
of the cluster redshift distribution. 
This is reassuring. Harder is the intepretation of the higher-redshift results. In this case 
clusters inhabited by point-like and sub-structured radio sources 
present a very similar dependence of the X-ray luminosity on the temperature. 
Whether this effect is due to the fact that the 'point-like' sample of sources 
in this redshift range also includes a fraction of sub-structured (but unresolved) objects, 
therefore producing a net mix of the two populations or whether its explanation resides 
in the more physical effect that at high redshift we are preferentially selecting higher-mass 
systems where the absolute effect of extended radio-loud AGN on the ICM is smaller is 
difficult to tell. Higher-resolution radio observations of the cluster centres are needed to disentangle 
these two effects.

The average $L_X-kT$ behaviour obtained by considering both populations of extended 
and point-like radio sources is in remarkable agreement with what is found in the literature 
for the whole cluster population (e.g. White, Jones \& Forman 1997; Arnaud \& Evrard 1999; 
Popesso et al. 2005).  This result, together with the fact that we expect most X-ray 
selected clusters to host a central radio object brighter than $\sim 10^{20}$ W/Hz/sr 
(see \S 5.1), suggests that the observed $L_X\propto T^3$ scaling is due to the 
superposition of two separate populations: clusters inhabited by unresolved 
radio objects and clusters with extended radio emission.

The above analysis underlines the fundamental importance of extended
radio-loud AGN in breaking the self-similarity of clusters, at least on scales
$kT\simlt 3$ KeV.  At this point one can wonder how the level of radio
activity depends on the cluster temperature. A plot of $L_R$ vs $kT$ is
presented in the right-hand panel of Figure \ref{fig:LvsT}. Independent of the
radio morphology, there is a much greater spread than in the $L_X$ vs $kT$
plot. However, for $kT\simlt 3$ keV, i.e. in the regime where radio-loud AGN
heating is very efficient, the $L_R-kT$ relation is relatively tight,
providing some evidence -- at least in low-mass systems -- for a relation
between radio luminosity and cluster temperature, in that hotter clusters host
more powerful radio sources. Note that this result is found to hold for both
unresolved and sub-structured radio sources.

\begin{figure}
%\vspace{8cm}  % amount of vertical space needed
\includegraphics[width=8.0cm]{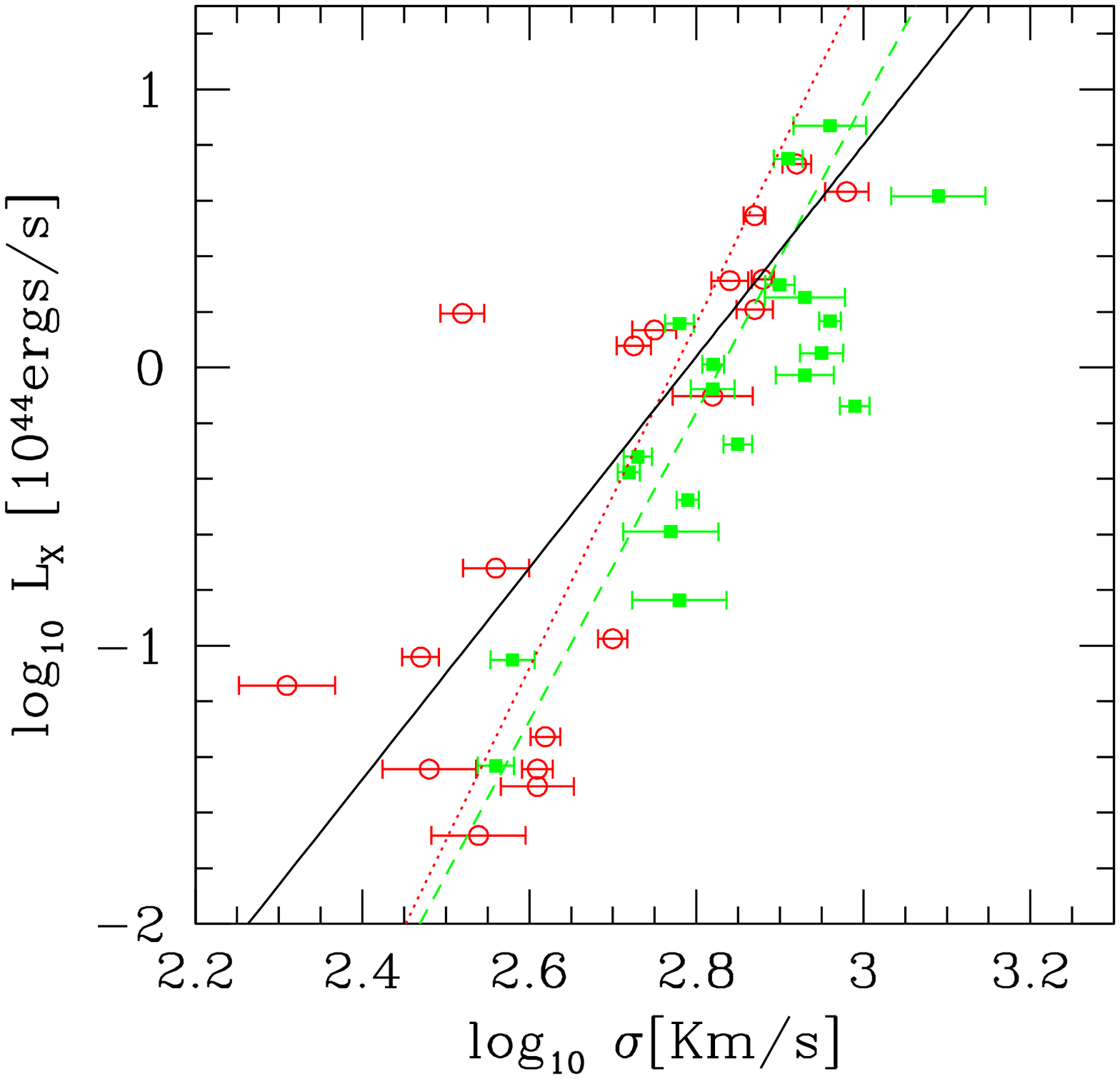}
%\special{psfile=lmechvssigma.ps hoffset=-10 voffset=-80 vscale=44
%hscale=44 angle=0} 
\caption{Ratio between mechanical and X-ray luminosity versus cluster velocity dispersion.
Open (red) circles identify 
point-like radio sources while filled (green) squares represent extended ones. The 
(green) dashed and (red) dotted 
lines respectively indicate the $L_{\rm mech}/L_X-\sigma$ best fits to the sub-populations 
of extended and point-like radio sources (see Table 1); the solid line is the best fit 
to the whole radio cluster sample.
\label{fig:lmech}}
\end{figure}

\subsection{Velocity Dispersions}
 
Velocity dispersion measurements for this part of our analysis have been taken
from the VIZIER database and in most cases rely on the works of Mahdavi et al.
(2000) and Mahdavi \& Geller (2001).  X-ray luminosities have been compared to
velocity dispersions for the 39 clusters for which measurements were available
in the left-hand panel of Figure \ref{fig:Lvssigma}. Best-fit values for the
fits are provided in Table 1.  The trend seen for the whole population
of clusters inhabited by a central radio source ($L_X\propto \sigma^b$, with
$b=3.8\pm 0.4$) is in general agreement with what is found in the literature for
different sets of clusters and groups of galaxies (e.g. $b\sim 3.6$, Popesso
et al. 2005; $b\sim 4.4$, Madhavi \& Gheller 2001; $b\sim 3.9$ for the
clusters considered in Osmond \& Ponman 2004). This behaviour, together with what was
found in Section 5.2 for the $L_X-kT$ relation, show that
radio-identified clusters provide a fair sample of the entire cluster
population.

However, the similarity breaks if one considers clusters inhabited by
unresolved and extended radio sources separately. For the former case, we find
that $b=6.2^{+0.4}_{-0.3}$, while in the case of extended radio emission we
find $b=5.5^{+1.0}_{-0.8}$. The difference in the observed behaviour of these
two sub-groups with the general case (cfr.  Figure \ref{fig:Lvssigma}a) may be
attributed to the fact that in our sample extended radio sources seem to be 
systematically associated to higher velocity dispersion systems. Even though the available
data is not good enough to provide more than a $1\sigma$ evidence for this
effect, the net result of combining extended and point-like radio sources is a
flattening of the $L_X-\sigma$ relation to the 'standard' value $b\sim 4$. As
already observed in Section 5.2, it is intriguing that the sum of two populations
with dissimilar behaviours gives rise to a trend that matches previous
measurements obtained for different cluster sets. This suggests again
  that clusters can be divided into two rather different groups: clusters that
  host point-like radio sources and those associated with extended radio
  emission.

It would be interesting to investigate the $L_X-\sigma$ relation in the
low-mass/low-luminosity region ($L_X\le 10^{43}$ erg/s, $kT\simlt 3$ KeV)
where the discrepancy in the $L_X-kT$ behaviour between extended and
point-like sources becomes more pronounced.  Unfortunately, the available data
on velocity dispersions only probe the $L_X\simgt 10^{42.5}$ erg/s regime, so
that this is currently not possible.  For the same reason, we cannot detect
any difference in the relation between temperature and velocity dispersion
(right-hand panel of Figure \ref{fig:Lvssigma}), a part from a marginal
preference for extended radio sources to reside in higher velocity dispersion
systems.  The slope of the $\sigma$ vs $kT$ relation is $b\sim 0.5$ in all
cases, again in agreement with most of the results found in the literature.

By analysing the cavities and bubbles that are produced in clusters and groups by the 
interaction between radio sources and the surrounding hot gas, Birzan et al. 
(2004) provide a 
useful empirical relation to estimate the mechanical luminosity released by the central 
radio source into the ICM:
\begin{equation}
\frac{L_{\rm mech}}{10^{36} \rm W}=(3.0\pm 0.2)\left(\frac{L_{1.4\rm GHz}}{10^{25} \rm W 
Hz^{-1}}\right)^{0.40\pm 0.13} \label{eq:4}
\end{equation}
Even though the 1.4 GHz synchrotron luminosity is not too reliable at predicting 
the mechanical luminosity as can be seen from the large errors in the exponent of 
Eq. ~\ref{eq:4}, we will use it to calculate 
the energetic balance between AGN heating and radiative losses 
(as in Best et al. 2007 and Birzan et al. 2004).
This is presented in Figure \ref{fig:lmech} as a function of the cluster velocity 
dispersion, 
where once again solid squares are for extended radio sources, while open circles represent 
point-like structures. 
In agreement with Best et al. (2007) we find that, while on the smallest mass scales 
($\sigma\simlt 400$ Km/s) radio heating balances radiative losses, such a heating mechanism 
falls short by as much as a factor 100 for the most massive clusters in our
sample.
 
There is no significant difference between the behaviour of extended and
point-like sources ($L_{\rm mech}/L_X\propto \sigma^{-7}$), even though the
data does not allow us to conclude anything about the group regime.

Thus, for the smaller mass systems we have found the following: (i)
mechanical heat and radiative losses balance and (ii)
there is a correlation between X-ray luminosities and cluster temperatures
and between radio luminosities and
cluster temperatures in the low temperature regime  (Section 5.2).  This
suggests the conclusion that in objects such as groups or small clusters of
galaxies there is a strong interplay between central radio source and the surrounding gas.
In this case, more powerful radio sources lead to hotter systems. The
eventual presence of extended radio
structures such as jets and lobes is of fundamental importance to the
thermal state of the ICM, as they can carry energy throughout the whole 
cluster and appear to be more efficient at heating the surrounding medium. 
The heating provided by the central radio
source is also found to be sufficient to balance radiative losses.

The situation is different at higher masses. In this case, we find that there 
is no correlation between the radio luminosity of the central source and the cluster 
temperature. Also, the effect of the radio-loud AGN on the thermal state of the 
cluster is independent of the capabilities of such an object to permeate 
the surrounding medium (no difference in the $L_X-kT$ relation between extended 
and point-like sources). AGN heating is found to be insufficient to counterbalance the 
radiative losses of the ICM, with a relative energetic importance which 
steeply decreases (as $\propto \sigma^{-7}$) with cluster mass. 

A different source of prominent heating must be invoked in this latter case.
One such mechanism may be thermal conduction (see e.g.  Br\"uggen et al. 2003;
Hoeft \& Br\"uggen 2004; Roychowdury et al. 2005; Ruszkowski \& Begelman 2002; 
Fujita \& Suzuki 2005; Narayan \& Medvev 2001; Voigt \& Fabian 2004).
Hoeft \& Br\"uggen (2004) have provided evidence for the relative importance
of heating due to thermal conduction to increase with cluster mass, in
agreement with our results.

\subsection{Cooling core clusters}

\begin{figure}
%\vspace{8cm}  % amount of vertical space needed
\includegraphics[width=8.0cm]{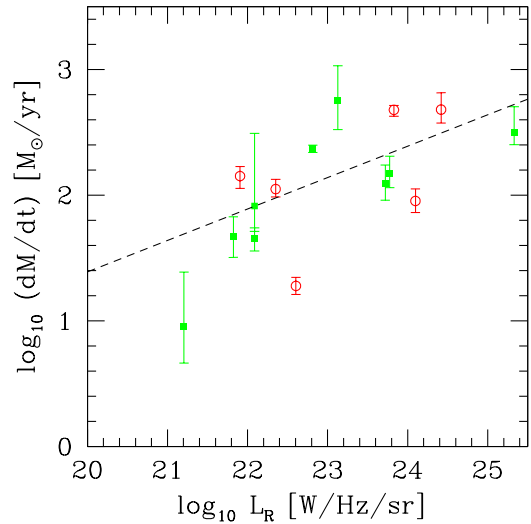}
%\special{psfile=Lrvscool.ps hoffset=-10 voffset=-80 vscale=44
%hscale=44 angle=0}
\caption{ X-ray luminosity within the cooling volume versus radio luminosity $L_R$ of 
the central source. Symbols are as before. 
\label{fig:cool}}
\end{figure}

Clusters of galaxies can be separated into two classes: clusters with dense gaseous cores in which 
the cooling time is less than the Hubble time, (cool core clusters), and cluster with less 
dense gaseous cores (non-cool core clusters).

Chen  et al. (2007) have investigated the influence of cool cores on cluster scaling relations. 
Their analysis has shown that in cool core clusters the X-ray luminosity is enhanced over that of 
non-cool core clusters, while other parameters such as temperature, mass and gas mass are less 
affected 
by the occurrence of a cooling core. This has been explained by the fact that at least some of the 
non-cool core clusters are in dynamically young states compared with cooling core clusters. 
Alternatively, 
the non-cool core clusters might have had their cool cores heated by a radio-loud AGN. This is 
suggested by Fig.~\ref{fig:LvsT}a which shows the $L_X-T$ relation for clusters with an extended 
radio source and for those without it. Our result indicates that the scaling relations for clusters 
are significantly affected by the presence of a radio-loud AGN at their centres.

In the work by Birzan et al. (2004), a close correlation between the mechanical 
energy output of the jets and the energy loss of the central X-ray plasma by cooling (related 
to the traditionally determined cooling flow mass deposition rate) has been established. This 
lends support to the idea that the feeding of the AGN is connected to the cooling 
rate and that the energy output of the AGN regulates the cooling core structure by feedback 
processes. In any model of feedback, the power of the central AGN should be regulated by the supply 
of fuel from the ICM. Thus, as in a thermostat, large radiative losses should give rise to
a strong, central radio source, which in turn will heat up the ICM (see Churazov et al. 2003). 
The more strongly a cluster cools at its centre, the more fuel can be supplied to the central 
AGN which, in turn, can drive more energetic radio lobes through the ICM which gets heated until 
the central cooling time goes up, the mass deposition rate goes down and the supply of fuel is 
shortened.

Cooling flow measurements for 15 clusters belonging to our sample have been taken from 
White et al. (1997). 
Figure \ref{fig:cool} shows the dependence of the radio luminosity of the central source, 
$L_R$, on the cooling luminosity $\frac{dM}{dt}$, i.e. the X-ray luminosity within the cooling 
volume (we note that, even though White et al. 1997 quote their values in an $\Omega_0=1$, 
$h_0=0.5$ cosmology, converting these measurements to the adopted $\Lambda$CDM cosmology would 
merely correspond to a vertical shift of the $\frac{dM}{dt}$ vs $L_R$ relation which does not 
affect our analysis and conclusions).
Given the paucity of sources and the fact that they did not show any 
different behaviour, in this case the best-fit value for  
 $\frac{dM}{dt}$ vs $L_R$ is only given for the whole sample of clusters which host a central radio 
source, independent of whether this source presents an extended or a point-like structure.
The plot of the classical mass deposition rate versus the radio luminosity shows a clear 
correlation 
($\frac{dM}{dt}\propto L_R^b$ with $b=0.25\pm 0.02$ cfr. Table 1) which is in agreement with 
feedback models (e.g. Churazov et al. 2003) for the fueling of cluster radio sources. 

\section{Clusters with central radio emission: under-luminous or overheated?}

\begin{figure}
%\vspace{8cm}  % amount of vertical space needed
\includegraphics[width=8.0cm]{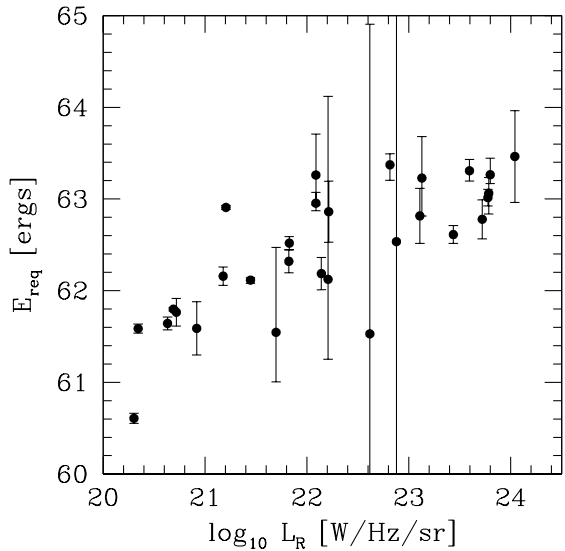}
%\special{psfile=requiredenergy.ps hoffset=0 voffset=-70 vscale=40
%hscale=40 angle=0} 
\caption{Thermal energy required to heat the clusters which host in their centres 
extended radio sources from the predicted to the measured temperature 
(cfr. Figure 7a) as a function of radio power (see text for details).    
\label{fig:reqenergy}}
\end{figure}

\begin{figure*}
%\vspace{8cm}  % amount of vertical space needed
\includegraphics[width=8.0cm]{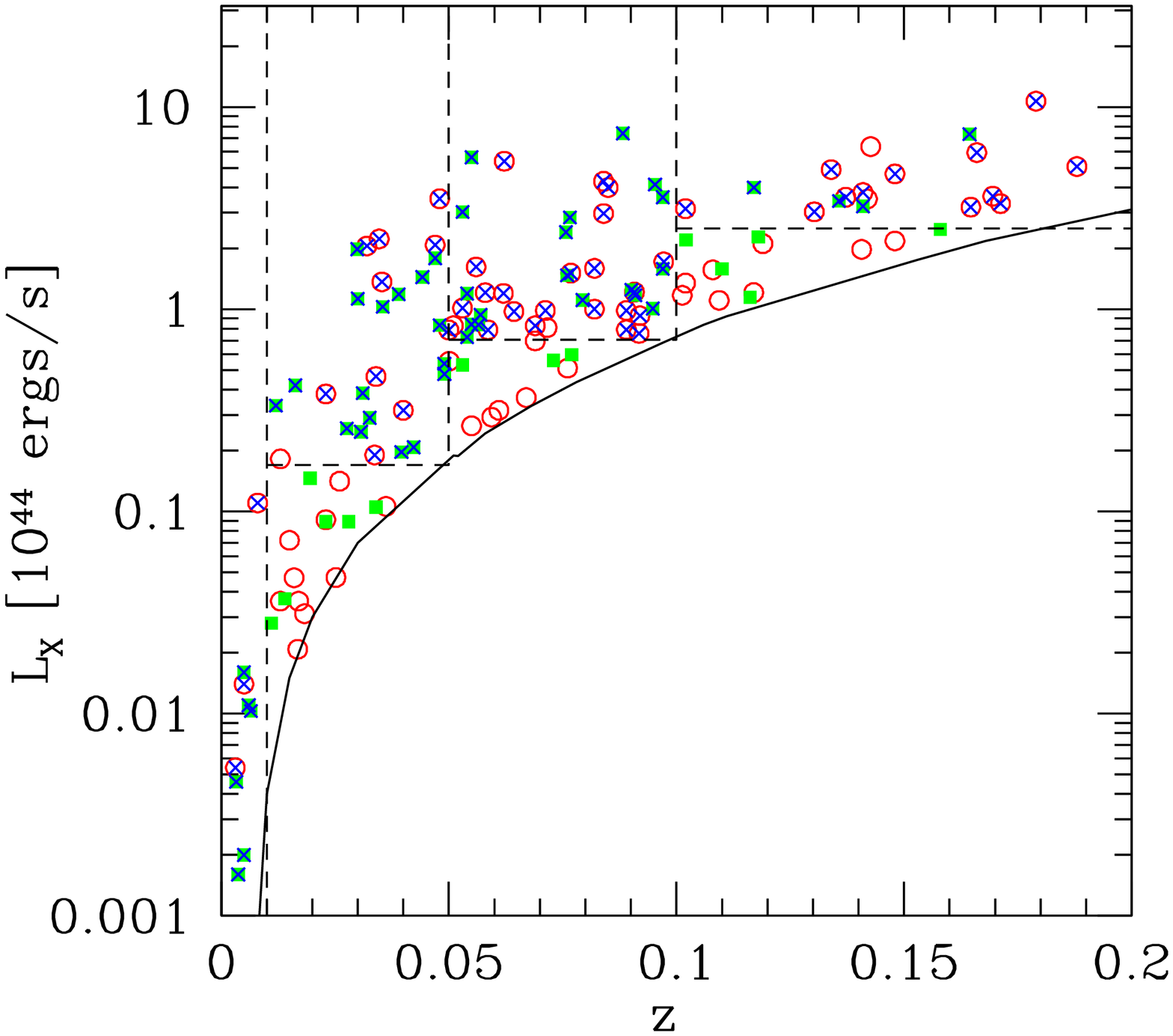}
\includegraphics[width=8.0cm]{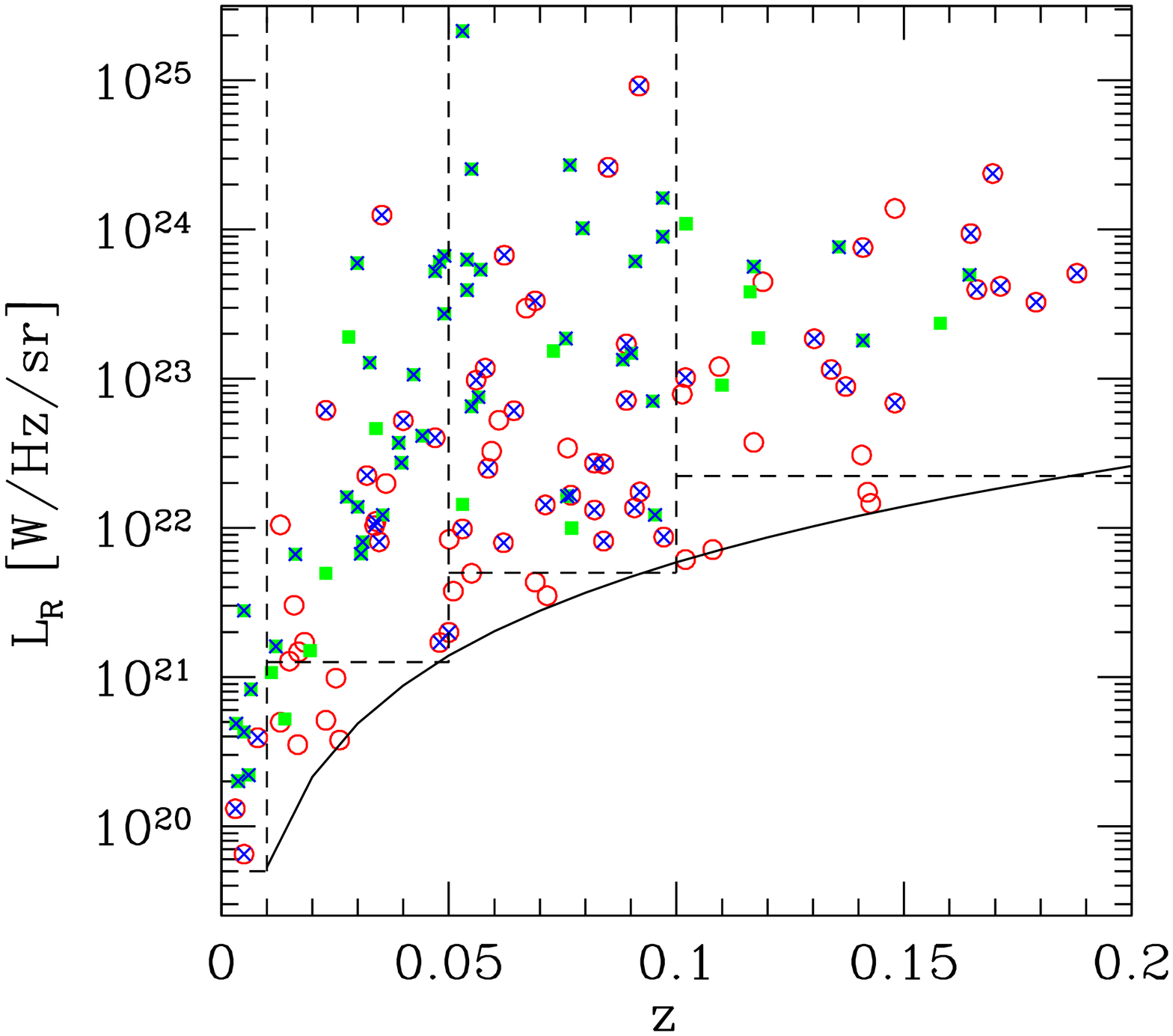}
%\special{psfile=zlimits_x.ps hoffset=-10 voffset=-80 vscale=44
%hscale=44 angle=0} 
%\special{psfile=zlimits_radio.ps hoffset=250 voffset=-80 vscale=44
%hscale=44 angle=0}
\caption{Distribution of X-ray (left-hand panel) and radio (right-hand panel) 
luminosities as a function of redshift for the 145 NORAS+REFLEX clusters 
endowed with a central radio-active component. The solid lines show 
the minimum luminosities probed by the X-ray and 1.4~GHz surveys 
corresponding respectively to the limiting fluxes of 
$3\cdot 10^{12}$~erg s$^{-1}$ cm$^{-2}$  and 3~mJy. Open (red) circles correspond to 
point-like radio sources, while filled (green) squares identify sub-structured objects. 
Crosses highlight those systems (91, out of which 45 show signatures of  
of extended emission) which fulfill the luminosity requirements 
(\ref{eq:limits}) adopted for the Monte Carlo 
simulations (dashed horizontal lines both in the left-hand and right-hand 
panels) and that therefore have been included in the analysis presented 
in this section (see text for details).    
\label{fig:zlimits}}
\end{figure*}

In Section 5.2 we have seen how clusters that host a radio source endowed 
with an extended structure show a departure from the typical 
$L_X-kT$ relation found for the cluster population as a whole (cfr. Figure 
\ref{fig:LvsT}a), a departure that is particularly remarkable in the case of low-mass 
systems. What is not yet clear from our analysis is if extended radio sources 
cause an under-luminosity or an overheating of the associated cluster. 
If it is indeed a temperature increment caused by the central radio galaxy, 
one may expect the corresponding heat excess to be correlated with the radio power of the central 
source. 
%Note that a tentative evidence for this effect in the case of low-mass systems is shown in 
%Figure~6b.

To investigate this possibility, we have considered the quantity  
$E_{\rm req}=3Nk\Delta T/2=3M_{\rm gas}k\Delta T/2\mu m_{\rm H}$ which is the energy required to 
heat the ICM from the predicted to the measured temperature. In the above expression, 
$N$ is the total number of particles, $M_{\rm gas}$ the gas mass of the cluster and $\Delta T$ 
the temperature increment of clusters which host an extended radio source with respect to the 
 $L_X-kT$ relation obeyed by clusters with unresolved central sources. 
The gas mass was computed from the total mass of the cluster by using a constant gas mass fraction 
of 0.2. We have plotted the required energy as a function of radio power in 
Fig.~\ref{fig:reqenergy}. 
The radio luminosity and the heat input needed to produce the observed temperature increment 
in clusters hosting extended radio sources appear 
to be correlated, although with a large scatter. This correlation favours a model where the 
temperature increase is caused by radio galaxy-induced heating. As noted by Croston et al. (2005) 
who performed a similar analysis for groups of galaxies, the large scatter in this plot is not 
surprising given that there are many unknown factors such as the age of the source, its history 
and size.

The issue of temperature increment can be investigated further by performing 
Monte Carlo simulations to be compared with the X-ray and radio luminosity 
distributions of the present sample. Possible biases due to the 
joint effects of the X-ray and radio selection functions (highlighted by the solid 
curves in the two panels of Figure \ref{fig:zlimits}), have been removed by limiting 
our analysis to four contiguous redshift regions which are complete both in X-ray 
luminosity and in 1.4~GHz radio power. In order to maximize the number 
of sources available for statistical analyses, these four regions in 
the $z-L_X-L_R$ space have been identified as follows: 
\begin{eqnarray}
1)\;z\le 0.01; L_X\ge 0.0015\cdot 10^{44}; L_R\ge 10^{19.5};\;\;\;\;\;\;\;\;\;
\nonumber\\
2)\; 0.01<z\le 0.05; L_X\ge 0.17\cdot 10^{44}; L_R\ge 10^{21.1};\;\;\nonumber\\
3)\;0.05<z\le 0.1; L_X\ge 0.708\cdot 10^{44}; L_R\ge 10^{21.7};\;\;\nonumber\\
4)\;0.1<z\le 0.2; L_X\ge 2.51\cdot 10^{44}; L_R\ge 10^{22.35},\;\;\;\;\;
\label{eq:limits}
\end{eqnarray}
-- where X-ray luminosities are in erg/s and radio luminosities 
in W/Hz/sr -- and are those enclosed within the dashed lines 
in Figure \ref{fig:zlimits}. 91 clusters in our dataset fulfill the above requirements 
(crosses in Figure \ref{fig:zlimits}), out of which 45 host sub-structured/extended 
radio sources (filled green squares with crosses on top in Figure \ref{fig:zlimits}).

\begin{figure*}
%\vspace{8cm}  % amount of vertical space needed
\includegraphics[width=8.0cm]{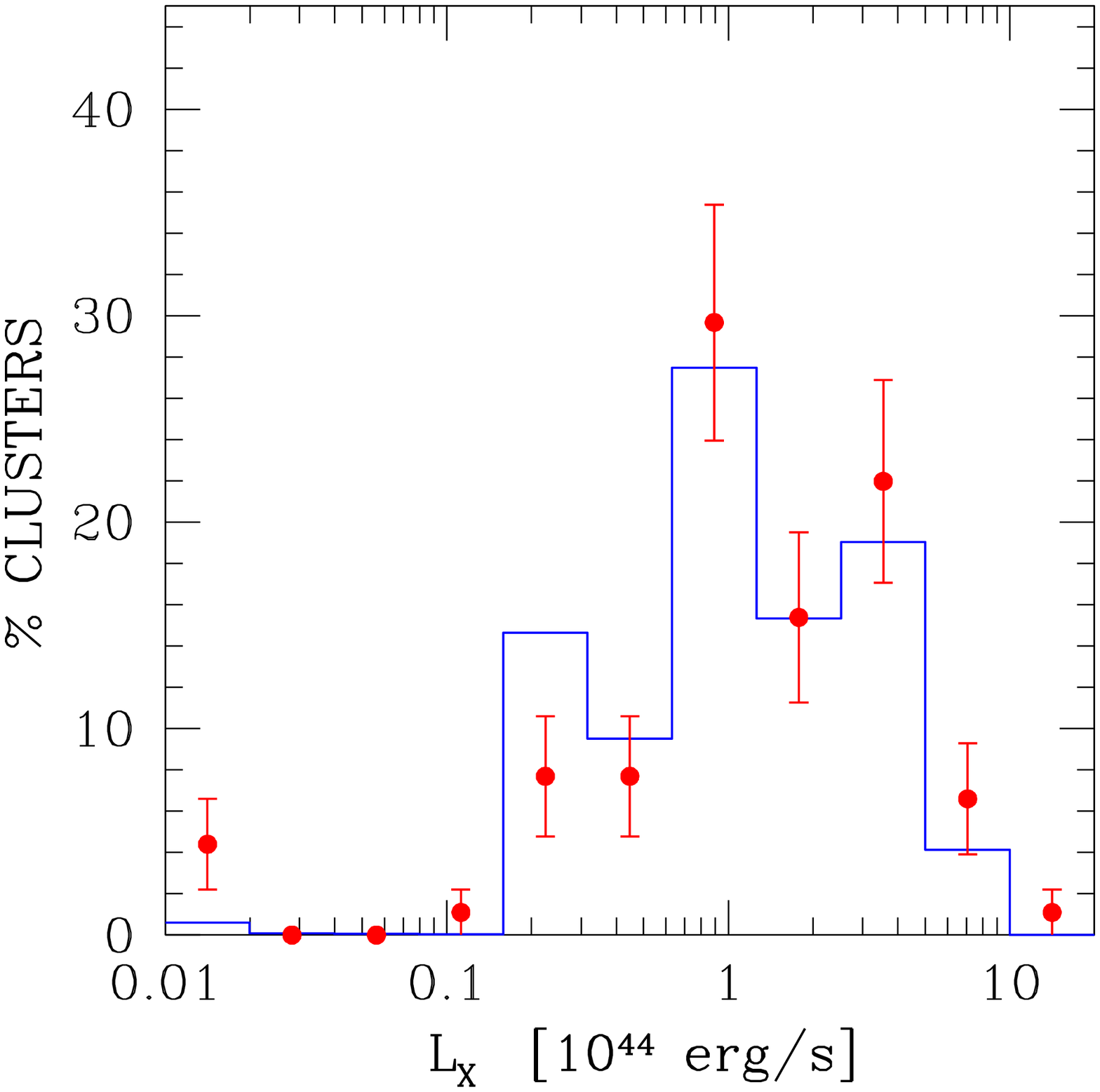}
\includegraphics[width=8.0cm]{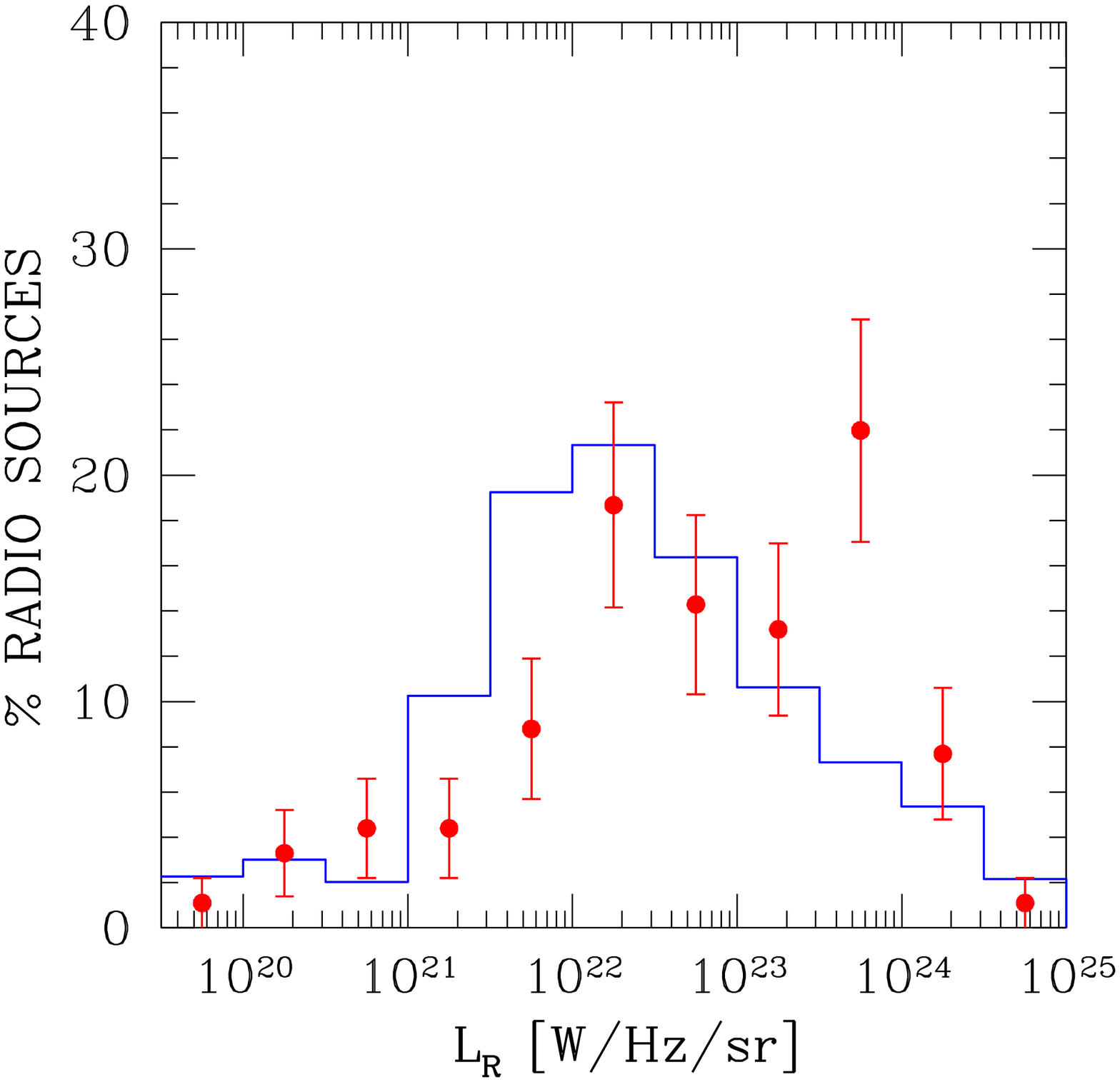}
%\special{psfile=hist_x_power.ps hoffset=-10 voffset=-80 vscale=44
%hscale=44 angle=0} 
%\special{psfile=hist_radio_power.ps hoffset=250 voffset=-80 vscale=44
%hscale=44 angle=0}
\caption{Fractional luminosity distributions of X-ray selected clusters 
inhabited by a central radio-active AGN. The solid points with associated 
errorbars correspond to the 91 sources extracted from the sample  
presented in this work by requiring the conditions (\ref{eq:limits}) 
to be fulfilled. 
Solid lines represent the results of the Monte Carlo simulations. 
\label{fig:hist_power_all}}
\end{figure*}

\begin{figure*}
%\vspace{8cm}  % amount of vertical space needed
\includegraphics[width=8.0cm]{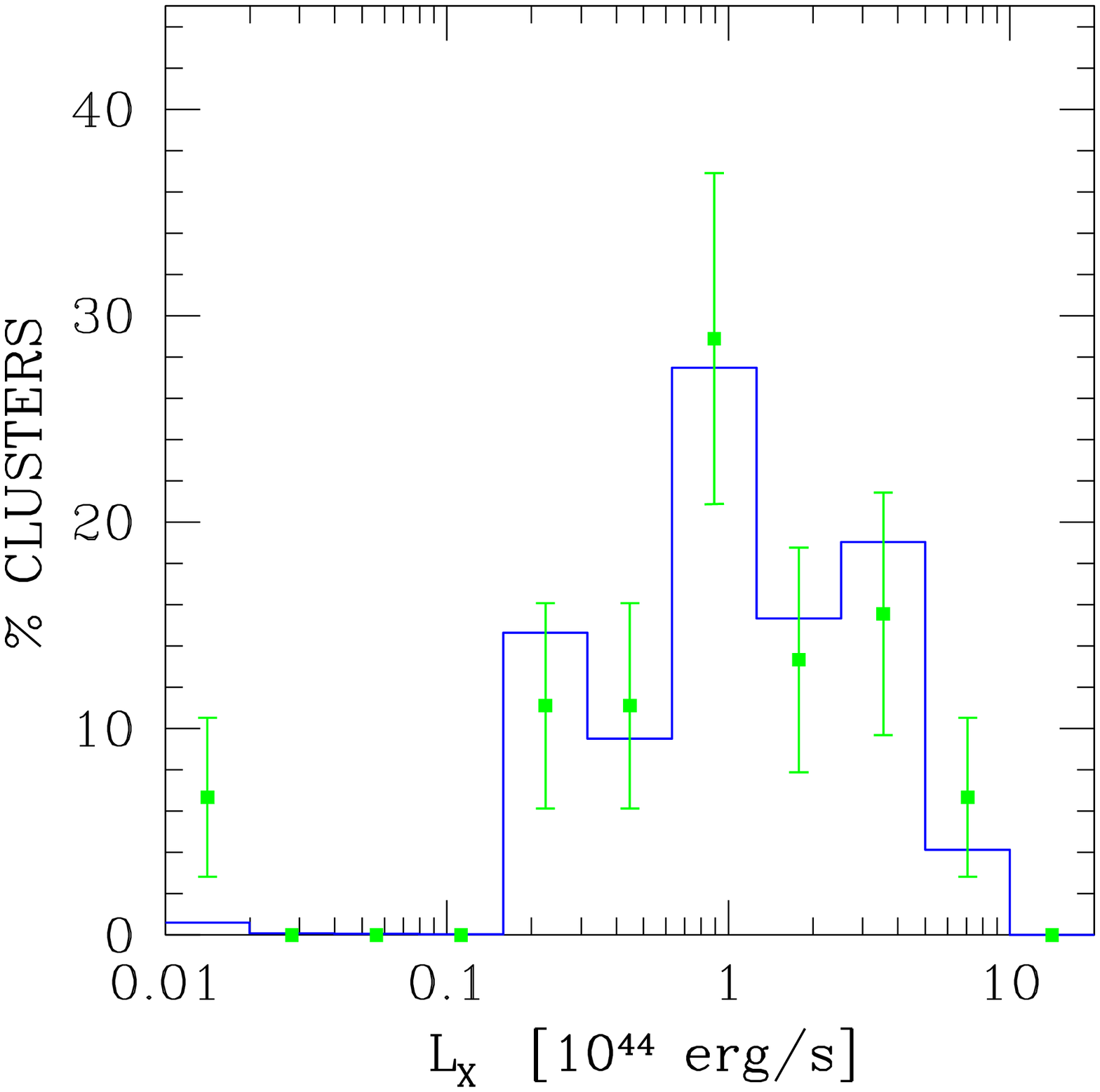}
\includegraphics[width=8.0cm]{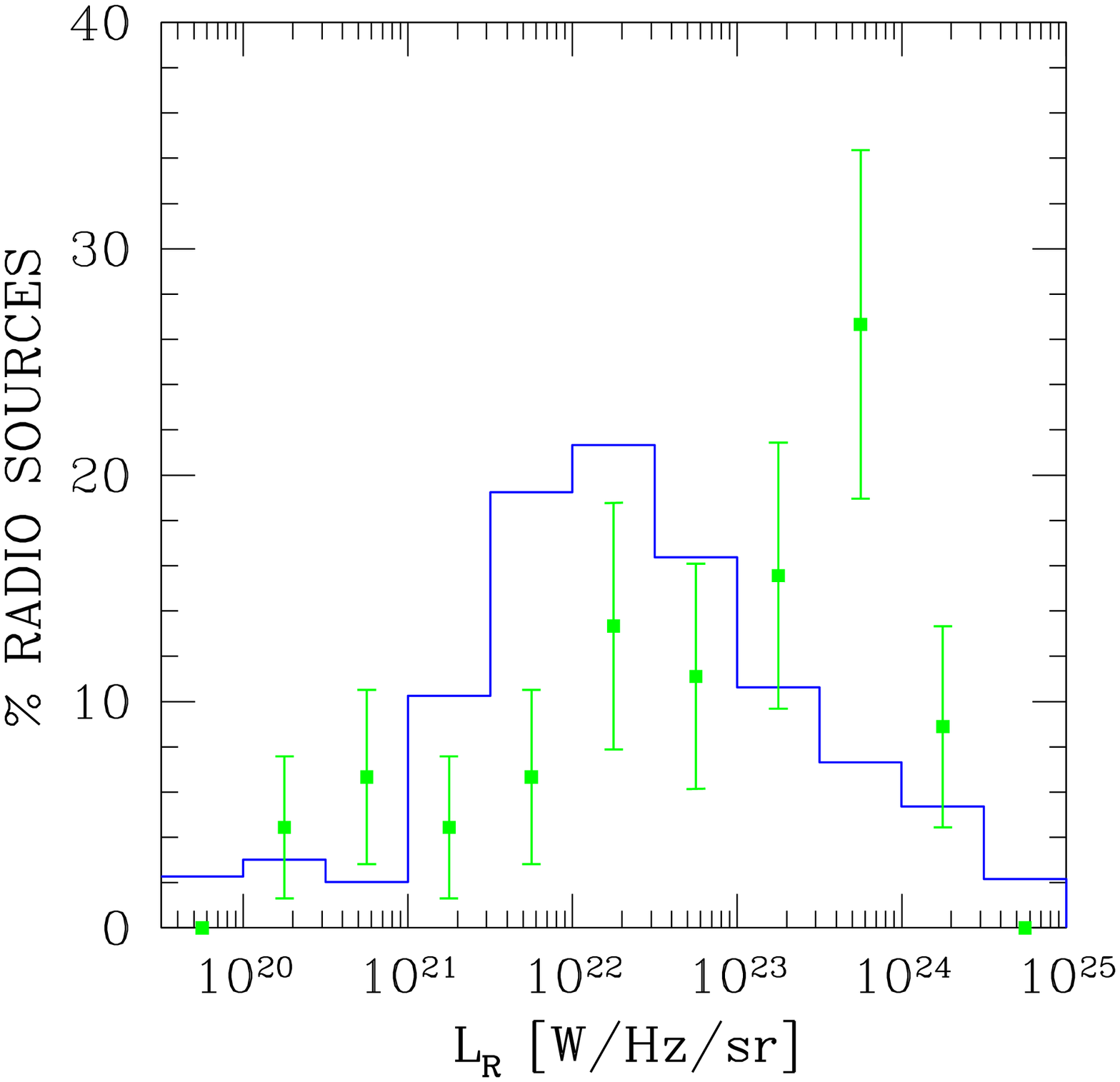}
%\special{psfile=hist_x_power.ps hoffset=-10 voffset=-8
%\special{psfile=hist_x_power_struct.ps hoffset=-10 voffset=-80 vscale=44
%hscale=44 angle=0} 
%\special{psfile=hist_radio_power_struct.ps hoffset=250 voffset=-80 vscale=44
%hscale=44 angle=0}
\caption{The same as in Figure~\ref{fig:hist_power_all} but for 
the sub-population of extended radio sources (45 objects).
\label{fig:hist_power_struct}}
\end{figure*}

Simulated samples in the $z-L_X-L_R$ space have been obtained by assuming 
that the cluster X-ray luminosity and radio luminosity of the central source 
are not correlated with each other. 
Each simulated object in the four regions considered in our analysis 
was then given a triplet of values $z$, $L_X$ and 
$L_R$, where the X-ray luminosity was 
assigned according to the B\"ohringer at al. (2002) cluster luminosity function 
as derived from the REFLEX cluster survey once their sample was corrected  
both for missing flux and flux error:
\begin{eqnarray}
\Phi(L_X)dL_X=\Phi_0\;\left(\frac{L_X}
{L_*}\right)^{-\alpha}{\rm exp}\;\left(-\frac{L_X}{L_*}\right)\;\frac{dL_X}{L_*},
\end{eqnarray}
with $\alpha=1.69$, $\Phi_0=1.07\cdot 10^{-7}h^3_{0,50}$~Mpc$^{-3}$ and 
$L_*=8.36\cdot 10^{44}h_{0,50}^{-2}$~erg/s, where $h_{0,50}=h_0/0.5$.\\

Radio luminosities were instead drawn from the Dunlop \& Peacock (1990) 
radio luminosity function for steep-spectrum sources (model 7 for pure 
luminosity evolution), proved to provide 
a very good fit to the local, AGN-fuelled, radio population 
(see e.g. Magliocchetti et al. 2002):
\begin{eqnarray}
\Psi(L_R,z)=\Psi_0\left[\left(\frac{L_R}{Lc(z)}\right)^\alpha+\left(\frac{L_R}{L_c(z)}
\right)^\beta\right]^{-1},
\label{eq:DP}
\end{eqnarray}
where $\alpha=0.69$, $\beta=2.17$, $\Psi_0=10^{-6.97}$ Mpc$^{-3}$ $(\Delta {\rm log}_{10} 
L_R)^{-1}$ 
and where the evolving break luminosity 
$L_c(z)$ can be expressed as ${\rm log}_{10}L_c(z)=26.22+1.26 z-0.26 z^2$ (in 
W$^{-1}$ Hz$^{-1}$ units). 
Since the above expression was derived for an $\Omega_0=1, h_0=0.5$ (EdS) universe, following 
Dunlop \& Peacock (1990) we have rewritten equation (\ref{eq:DP}) in the concordance 
$\Lambda$CDM cosmology by using the relation:
\begin{eqnarray}
\Psi_1(L_{R,1},z) \frac{dV_1}{dz}=\Psi_2 (L_{R,2},z) \frac{dV_2}{dz},
\end{eqnarray}
where 1 and 2 refer to the old and new cosmology and $dV/dz$ is the volume element.\\
We performed 1000 Monte Carlo runs, where each run was set so to give the same number 
of sources per redshift interval with 
luminosities above the limits expressed in (\ref{eq:limits}) as in the real dataset.

The luminosity distributions of X-ray-selected clusters 
inhabited by a radio-loud AGN are shown in Figures \ref{fig:hist_power_all} 
and \ref{fig:hist_power_struct} which 
provide the fractional number of clusters per X-ray (left-hand panels) and 
1.4 GHz luminosity (right-hand panels) for both real data (points with 
associated errorbars) and simulated sample (solid lines. Note that, due to 
the high number of realizations, these values are virtually error-free). 

Two important conclusions can be drawn from the distributions shown in Figures 
\ref{fig:hist_power_all} and \ref{fig:hist_power_struct} . 
First, the X-ray luminosity distribution of clusters with a central 
radio-active counterpart is virtually identical to that featured by the whole 
population of X-ray selected clusters brighter than the same flux limit. 
In other words, the presence of a central radio-loud component, 
however powerful, does not affect the cluster X-ray luminosity. 
This result is found regardless to whether the radio AGN shows an extended or 
point-like structure. 
If we then combine the above finding with what emerged from Sections 5.2 and 5.3,
we can conclude that -- mostly in the case of small mass systems -- 
the interaction of extended radio sources with the surrounding environment causes 
a remarkable {\it overheating} of the available gas, heating which produces 
the steepening of the $L_X-kT$ relation observed in Figure \ref{fig:LvsT}a.
The contribution of AGN heating from a central radio source is less important {\it i)} 
in the case of point-like sources which are not observed to permeate the ICM 
and {\it ii)} -- independent of the type of radio object -- in the case of high mass systems 
which require additional forms of energy to balance radiative loss.
%%%%%%%%%%%%%%%%%%%%%%%%%%%%%%%%%%%%%%%%%%%%%%%%%%%%%%%%%%%%%%%%%%%%%%%%%%%%%%%%%%%%%%%%%%%%%%%%%%%
%We note that similar conclusions on the relevance of radio-AGN heating 
%in groups were reached by Croston et al. (2005) who base their claim on a much smaller number 
%of objects and do not make distinctions between point-like and extended radio structures. 

This is in agreement with the results displayed in Figure~\ref{fig:reqenergy}. 
Physically, it means that the radio sources displace only a small fraction of 
the X-ray luminous ICM 
so that there is no noticeable luminosity deficit in clusters with radio structure. 
However, simulations show that if a significant fraction of the jet energy is transferred into 
potential energy thus leading to an expansion of the cluster atmosphere, also the luminosity of 
the ICM decreases (see Heinz et al. 2006). Such simulations also suggest that a 
significant fraction 
of the jet energy is thermalised in the ICM on relatively short time scales. 
On long time scales, however, the energy injected into the ICM by radio galaxies must end up 
mostly as potential energy by the virial theorem, thus implying long-term temperature increments to 
be small. This indicates that our analysis is unveiling the short-term effects of radio-loud 
AGN heating. The extended radio structures in fact originate from recent AGN activity and are 
found in correspondence with ongoing heating in the ICM. On the other hand, those clusters with no 
extended structures have not had a recent 
(i.e. longer than the life time of a radio lobe) outbreak of radio-loud AGN activity and therefore 
show no signs of recent temperature increments.

While the X-ray luminosity of a typical cluster remains unaffected by the 
presence of a central radio-emitting component, our simulations clearly show 
that this is not the fate for the 1.4~GHz 
luminosity of a radio galaxy set at the bottom of the potential well of a cluster. 
In fact, the right-hand panels of Figures \ref{fig:hist_power_all} and 
\ref{fig:hist_power_struct} highlight that there is a remarkable deficit of 
low-luminosity ($10^{21}~{\rm W/Hz/sr}\simlt L_R\simlt 10^{22}$~W/Hz/sr) 
radio sources when compared to the whole $F_R\ge 3$~mJy population. 
At the same time, radio-loud AGN within the cluster centres seem to favour 
luminosities $\sim 10^{24}$~W/Hz/sr ($\sim 3\sigma$ discrepancy from the 
results for the whole 1.4~GHz radio population). This different behaviour 
between the radio-luminosity distribution of real sources and of simulated 
data sets is found regardless of whether the central radio source presents 
an extended structure or not.
This result implies a difference between the luminosity function of 
radio sources sitting in the cluster centres and that of the whole radio 
population, in the sense that the former one is much flatter at all luminosities 
$L_R\simlt 10^{24}$ W/Hz/sr. Indeed, Figure 4 of Best et al. (2007) clearly shows 
the different behaviour of the luminosity functions as derived for their sample of 
BCGs and in the more general case of radio sources associated to SDSS galaxies of 
different mass, even though the authors dismiss the evidence as 'tentative'.

The fact that 
low-luminosity radio sources are underrepresented in clusters
is most likely due to the 'strangling' effect caused by the overdense central 
regions of massive systems on radio objects that are not powerful enough 
to expand throughout the surrounding medium. On the other hand, if we combine 
the results presented in this section with the analysis 
performed in Section 5.1, we can conclude that extended structures and high radio 
luminosities are more likely to occur in sources that are located in rich environments 
such as groups or clusters of galaxies. 
This is likely caused by a tight interplay between the intracluster medium 
and the energy released by a central radio source, 
%which is powerful enough to beat the 'core barrier', 
which boosts the radio luminosity of the latter one.
%within the surrounding environment.} \\
Indeed, in agreement with our results, Barthel \& Arnaud (1996) find that for a 
fixed jet kinetic power, radio 
luminosities can be higher in clusters due to the confining effect of the 
dense ICM which reduces losses due to adiabatic expansion in the radio lobes, 
therefore enhancing the radio synchrotron emission.

\section{Conclusions}

We have combined the REFLEX and NORAS cluster surveys with the NVSS dataset to provide a 
list of X-ray selected clusters brighter than $3\cdot 10^{-12}$~erg s$^{-1}$cm$^{-2}$  
which host in their centres (${\rm dist}\le 1.5 \% \:r_{\rm vir}$) a radio source brighter 
than 3 mJy. Out of 550 clusters set in the whole sky north of $\delta=-40^\circ$, 148 
systems (corresponding to $\sim 27$ per cent of the total sample) 
show signatures of central radio emission. We find that three systems were imaged 
both in REFLEX and NORAS in the overlapping region $0^\circ\simlt \delta\simlt 2.5^\circ$ 
between the two surveys. By then removing double identifications we end up 
with a sample of 145 clusters that host a central radio source with $F_{1.4 \rm GHz}\ge 3$ 
mJy. 
Visual investigations of radio maps show that 61 of the radio objects 
(i.e. $\sim 42$ per cent of the sample) associated with cluster centres possess extended 
radio emission in the form of elongated blobs, double or triple/multiple structures. 
Archival data (mainly taken from the BAX and VIZIER databases) 
have provided extra information such as temperatures and velocity dispersions 
for some of the systems in our sample.

The main conclusions of our work can be summarized as follows:
\begin{enumerate}
\item
Although the fraction of clusters that host a central radio source 
brighter than $L_R=10^{22}$~W/Hz/sr is approximately constant with cluster mass 
and equal to 20\% (in agreement with the results of 
Best et al. 2007), we find that 11 out of 12 (i.e. 92 per cent) local/low-mass systems 
present central radio emission with luminosities 
$L_R\simgt 10^{20}$~ W/Hz/sr. This suggests that, either, radio sources 
preferentially inhabit low-mass clusters, or -- more likely -- that by going 
deep enough in radio flux, most of X-ray selected clusters will be found 
to host a central radio-loud AGN.
\item
The luminosity-temperature relation of clusters that
host a central radio source follows that found for the whole cluster population 
($L_X\propto T^{2.8}$ -- e.g. White, Jones \& Forman 1997; 
Arnaud \& Evrard 1999; Popesso et al. 2005). However, there is a significant 
discrepancy in the $L_X$ vs $kT$ relation between unresolved and extended radio sources. 
The latter population shows the much steeper trend $L_X\propto T^4$, while for 
unresolved structures one finds $L_X\propto T^{2.3}$, which is close to the self-similar result. The 
difference between the $L_X$ vs $kT$ trend for point-like and extended sources 
becomes particularly noticeable for temperatures $kT\simlt 3$~KeV, where almost 
all systems with extended radio emission lie below those associated with 
unresolved radio structures.
\item
The radio luminosities of the central sources show a steep ($L_R \propto T^6$) 
dependence on the cluster temperature for $kT\simlt 3$ KeV. Such a correlation 
is lost in more massive systems.
\item The X-ray luminosity vs velocity dispersion and the temperature vs 
$\sigma$ relations of our sample are in agreement with those found in the literature 
($L_X \propto \sigma^{3.8}$; $kT \propto \sigma^{2}$; e.g. Popesso et al. 2005; 
Madhavi \& Gheller 2001; Osmond \& Ponman 2004). No 
significant differences are found between point-like and extended 
sources, except for a mild preference for the latter population to appear in 
higher velocity systems. We remark though that the available 
$\sigma$ measurements do not allow us to probe the very low-mass regime where 
the different behaviour between point-like and extended sources 
could be more prominent.
\item  
The mechanical luminosity provided by the central radio source is found to balance 
radiative losses only in small (i.e. $\sigma\simlt 400$ km/s) systems.
\item
It seems plausible that heating by low-power (FR I) radio galaxies can explain the absence 
of cooling flows in clusters of galaxies. This is in agreement with conclusions by various other 
authors (e.g. Fabian et al. 2003; Croston et al. 2005), who concentrated their analysis on 
groups of galaxies. However, for the largest clusters, the energy 
released by the radio sources may not be sufficient to balance radiative losses.
\item
Monte Carlo simulations show that the X-ray luminosity of a cluster is not affected 
by the presence of a central radio source, however powerful. The luminosity 
distribution of the present sample of radio-identified systems is the same as 
that of the whole cluster population. 
Combining this result with what we found in point (2), we can conclude 
that the presence of radio sources with an extended structure is responsible for the 
{\it over-heating} of the intracluster gas. The importance of such a heating dramatically 
increases in low-mass systems.
\item 
The luminosity distribution of radio galaxies sitting in cluster centres 
is very different from that of the total radio population. In fact, in the former 
case we find that low luminosities ($L_R\sim 10^{21}-10^{22}$ W/Hz/sr) are depressed, 
while higher luminosities are strongly boosted. The net effect 
on the radio luminosity function is a flattening in the whole luminosity range 
$L_R\simlt 10^{24}$~W/Hz/sr.
\end{enumerate}

Our results support a strong interaction between AGN radio emission and the ICM.
Radio sources are present in the centres of almost all local/small mass 
systems and we expect the majority of clusters to exhibit central radio 
emission with powers $L_R\simgt 10^{20}$ W/Hz/sr. 
The ICM strongly affects the luminosity of the radio source that sits in the cluster centre: 
low-luminosity objects are 'strangled' by the overdense central medium, while bright ones have 
their luminosities boosted by the interaction with the surrounding gas 
(see e.g. Barthel \& Arnaud 1996). 
Moreover, the existence of a central radio source -- 
especially if it exhibits an extended structure and resides in 
small-mass systems -- leads to a significant heating of the 
ICM.

Furthermore, radio-loud AGN are observed to 
provide enough mechanical energy to balance radiative losses. 
The effect of AGN heating from a central radio source becomes less important 
{\it i)} in the case of point-like sources which are not observed to permeate the ICM 
and {\it ii)} -- independent of the type of radio object -- for high mass systems. 
Clusters hotter than $kT\simlt 3$ KeV, do not obey the tight $L_R$ vs $kT$ 
relation observed for small systems, the effect on the thermal state of the cluster 
as provoked by the presence of a radio-active AGN is independent of the capabilities 
of such an object to permeate the surrounding medium (no difference in the $L_X-kT$ 
relation between extended and point-like sources) and AGN heating is found to be 
insufficient to balance radiative losses in the cluster.
In this case, a different source of heating such as thermal conduction 
(e.g. Narayan \& Medvev 2001; Voigt \& Fabian 2004) has to be invoked.
 
The data suggest a smooth transition between the radio-AGN heating mode 
and the thermal conduction mode.
Indeed, double heating models -- first developed by Ruszkowski \& Begelman (2002) --
have been proved to provide a good agreement with the observed cluster properties (see e.g. 
Br\"uggen et al. 2003; Hoeft \& Br\"uggen 2004; Roychowdury et al. 2005; 
Fujita \& Suzuki 2005). Within this framework, Hoeft \& Br\"uggen (2004) have shown  
the relative importance of heating due to thermal conduction to increase 
with cluster mass, in excellent agreement with our findings.

Obviously, it would be desirable to push 
this kind of analysis to fainter fluxes 
(both radio and X-ray) and higher redshifts to investigate whether some of our findings 
only hold in the local universe or are more general properties of the cluster population.
Forthcoming surveys like COSMOS (see e.g. Finoguenov et al. 2007) are expected to 
provide such answers.\\
\\

\noindent
{\bf Acknowledgments}\\
MM wishes to thank S.Borgani, A.Merloni, P.Rosati \& G. De Zotti for 
discussions and clarifications which greatly helped shaping up this work.   
MB wishes to acknowledge the support by the DFG grant BR 2026/4 within the Priority Program
``Witnesses of Cosmic History'' and the supercomputing grants NIC 1927
and 1658 at the John-Neumann Institut at the Forschungszentrum
J\"ulich. We thank the referee for helpful comments.

\end{document}